\documentclass[aps,prb,twocolumn,groupedaddress,showpacs]{revtex4-1}
\usepackage{graphicx}
\usepackage{bm}

\begin{document}


\title{Dynamical splitting of cubic crystal field levels in rare-earth cage compounds}


\author{Mehdi Amara}
\email{mehdi.amara@neel.cnrs.fr}
\affiliation{Univ. Grenoble Alpes, CNRS, Grenoble INP\footnote{Institute of Engineering Univ. Grenoble Alpes.}, Institut N\'eel, 38000 Grenoble, France}


\date{\today}

\begin{abstract}
The Crystalline Electric Field (CEF) influence is usually described by considering an ideally symmetrical rare-earth site. In the case of cage compounds, ample excursions of the rare-earth inside the cage require an adapted CEF description. A corrective, position dependent, CEF term accounts for the deviation from the perfect symmetry. In the paramagnetic range, a CEF level with orbital degeneracy thus acquires a width reflecting the rare-earth spatial distribution. In the case, frequent in cubic systems, of an orbitally degenerate CEF ground state at the center, this width introduces an additional energy scale, influential at low temperature. A spherical simplification allows to identify the major consequences of a cage-split ground multiplet: a Schottky-like anomaly appears in the specific heat with associated reduction of the magnetic entropy and alteration of the magnetic properties. Concomitantly, a centrifugal Jahn-Teller effect develops that expands the distribution of the magnetic ion and softens the rattling phonons. These effects are confronted with anomalous paramagnetic properties of rare-earth cage compounds, notably rare-earth filled skutterudites and hexaborides.
\end{abstract}


\maketitle


\section{Introduction}
Impulsed by an interest in thermoelectric applications, the investigation of metallic cage compounds has soared during the last decade. In these systems, atoms are enclosed in oversized cages, allowing relatively large excursions from their average positions. In some crystallographic structures, the cage can accommodate a rare-earth ion, giving rise to specific magnetic properties. The most investigated rare-earth cage compounds are filled skutterudites, that crystallize according to the LaFe$_4$P$_{12}$-type structure \cite{Jeitschko1977}. These compounds display a variety of intriguing features, as the heavy fermion and superconductor PrOs$_4$Sb$_{12}$ \cite{Bauer2002}, or the metal-insulator transition (MI) in PrRu$_4$P$_{12}$ \cite{Iwasa2005}, the non-magnetic ordering of PrFe$_4$P$_{12}$ \cite{Keller2001}, etc.. These unconventional behaviors echo those of an extensively investigated, but still elusive, series of rare-earth cage compounds: the rare-earth hexaborides. Among them, the most enigmatic CeB$_6$, which features a non-conventional ordering \cite{Effantin1985, Amara2012}.
\begin{figure}
\includegraphics[width=\columnwidth]{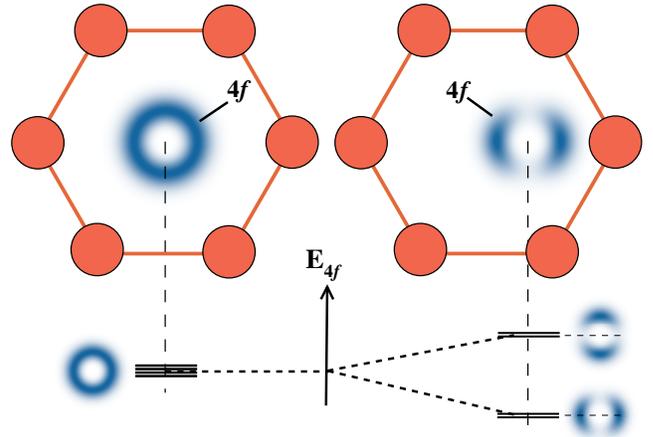}
\caption{\label{CagDiag} Schematic illustrating the effect, inside a high-symmetry cage, of an off-center position for a rare-earth ion (R$^{3+}$) with a non-Kramers CEF ground level. Outside the cage center, the non-Kramers level is split (energy scale E$_{4f}$). At low temperature, the off-center position and its altered 4$f$ electrons distribution (schematized densities) are statistically favored.}
\end{figure}
\newline
To understand the properties of rare-earth compounds, accounting for the effect of the Crystalline Electric Field (CEF) is essential. The CEF reflects the anisotropic environment of the 4$f$ ion and lifts, at least partly, the orbital degeneracy of the 4$f$ shell. As a result, the 2$J$+1 degenerate ground state multiplet is decomposed to form the CEF scheme, according to the point symmetries of the rare-earth site: this is a notorious and early success in the application of group theory to quantum mechanics \cite{bethe1929}. In the temperature range where magnetic phenomena occur, this CEF scheme is considered as a stable feature, used as a starting point for any microscopic description of the paramagnetic or ordering properties.\newline
In rare-earth intermetallic compounds, the most effective means, for the experimental determination of the CEF scheme, is neutron spectroscopy. However, in many instances of rare-earth cage compounds, difficulties emerge at the stage of the neutron spectroscopy investigation, notably for rare-earth filled skutterudites. For instance, the inelastic spectra of PrRu$_4$P$_{12}$ are inconsistent with the $T_h$ symmetry of the Pr site below the MI transition, displaying more CEF transitions than allowed, and show a spectacular broadening of these excitations above \cite{Iwasa2005}. In PrFe$_4$P$_{12}$, well defined CEF excitations appear only in the ordered state \cite{Park2008}, while, in PrOs$_4$P$_{12}$, they vanish very rapidly with increasing the temperature \cite{Iwasa2009}. For some light rare-earth hexaborides, neutron diffraction and Raman scattering investigations where successfully used for determining the CEF scheme \cite{Zirngiebl1984, Loewenhaupt1986}. In CeB$_6$, PrB$_6$ and NdB$_6$, the CEF ground states are well separated from the first excited ones and identified as non-Kramers, i.e. carrying an orbital degeneracy larger than the minimum reachable under an electrostatic influence. Their respective degeneracies are 4 ($\Gamma_8$), 3 ($\Gamma_5$) and 4 ($\Gamma_8$). However, in all three cases, the low temperature specific heat measurements yield values of the paramagnetic entropy much lower than expected from the degeneracy of these CEF ground states\cite{Fujita1980, PEYSSON1986, Lee1970, westrum1964}. In the case of CeB$_6$, a value consistent with the $\Gamma_8$ quadruplet is reached for temperatures one order of magnitude higher that the ordering one. This is possibly related to a Raman scattering observation: at low temperature, the cubic CEF quadruplet ground state spreads over an energy range of about 30 K in CeB$_6$\cite{Zirngiebl1984, Loewenhaupt1985}. Similar entropy anomalies, are observed as well in rare-earth filled skutterudites \cite{Ho2005,Yuhasz2006,Cichorek2014}.\newline
These recurrent inconsistencies force to reconsider the effect of the CEF in the cage context. The group theory approach relies on a system with ideal point symmetry, whereas physical systems are necessarily imperfect in this regard, due, at least, to thermal excitations and zero point fluctuations. This issue should be most severe in cage compounds. Indeed, how relevant is an approach based on the point symmetries at the cage center if the magnetic atom can substantially deviate from it? Moreover, in the here considered systems, the point symmetry at the cage center is high. At the center, the CEF ground state is thus likely to display an excess of orbital degeneracy. Such orbital degeneracies are known to cause Jahn-Teller instabilities: at low temperature, the system tends to spontaneously reduce its symmetry, simultaneously lifting the orbital degeneracy and reducing the electrostatic energy. Many instances of Jahn-Teller effect are found in systems where 3$d$ ions occupy sites of octahedral symmetry, such as in perovskite and spinel structures, in which they cause a distortion of the octahedra and, collectively, of the crystal \cite{VanVleck1939, Opik1957}. This cooperative kind of the Jahn-Teller effect is also found in rare-earth compounds \cite{Gehring1975}, notably in high symmetry insulators. These orbitally driven structural transitions are described considering a balance between the 4$f$ electrostatic energy and the elastic energy of the lattice.
In crystals where high symmetry cages accommodate loosely bound magnetic ions, the symmetry lowering doesn't require a distortion of the cage and, even less, of the crystal: it can be simply achieved with an offset magnetic ion (see Fig. \ref{CagDiag}). Accordingly, one can expect a Jahn-Teller effect to develop more easily in rare-earth cage systems than in conventional crystallographic structures.\newline
The work presented here attempts at answering these questions related to the specific CEF situation of cubic cage systems. The analysis is based on consideration of the lowest order correction to the CEF cubic hamiltonian for an offset ion. This results in a width for non-Kramers energy levels and in a specific, temperature dependent, 4$f$ electronic term in the cage potential well. 

\section{Model for an encaged 4$f$ ion}

\subsection{Definition of a cage system}
As a preliminary, an objective definition of the concept of rare-earth cage-compound is required. In these systems, the crystal structure is supposed to leave some latitude for the displacement of the guest ion inside the cage. One should then expect large amplitude vibrational modes (the so-called rattling), weakly coupled with the rest of the crystal. In order to identify a system as relevant to a cage compound approach, beyond an analysis based on comparisons involving ionic radii and interatomic distances, one can define tangible experimental criteria:\newline
- the Debye-Waller factor,\newline
- the phonons dispersion curves,\newline
which are below examined.

\subsubsection{The Debye-Waller factor}
As regards the displacement latitude, one direct information comes from the isotropic mean-square displacement of the rare-earth $U_{iso}$, involved in the Debye-Waller factor. This information can be derived from neutron or X-ray diffraction. Some data can be found in the literature for the RB$_6$\cite{CHERNYSHOV1997,Takahashi1999} and filled skutterudites from the antimonide series: RFe$_4$Sb$_{12}$\cite{Schnelle2008} and ROs$_4$Sb$_{12}$\cite{Kaneko2006,Yamaura2011}. At room temperature, $U_{iso}$ is about 0.005 \AA$^2$ in light rare-earths hexaborides and between 0.02 and 0.04 \AA$^2$ in the filled skutterudites, which yields a room temperature r.m.s. amplitude of the displacement $\sigma$ of about 0.07 {\AA} in the RB$_6$ and ranging between 0.1 and 0.2 {\AA} for the filled skutterudites.
In both series, the values $U_{iso}$ for the rare-earth guest is typically one order of magnitude larger than the mean-square displacements for atoms from the cage. This gives a quantitative credit to the rattling picture in these series.

\subsubsection{The phonons dispersion}
\label{Dispersion}
\begin{figure}
\includegraphics[width=\columnwidth]{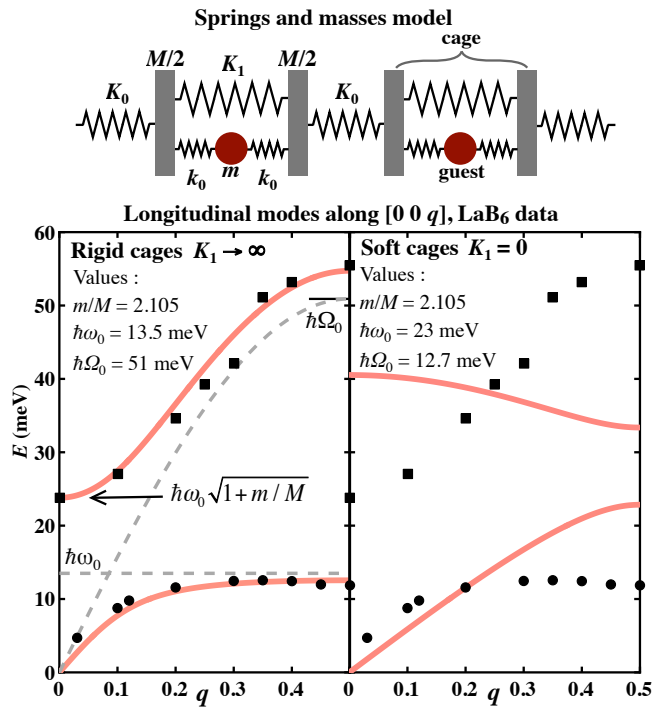}
\caption{\label{CageDisp} Upper part, a classical chain of springs (stiffnesses $k_0$,  $K_0$) and masses ($m$ for the guest and $M$ for the cage) for describing longitudinal modes in cage systems, including a cage stiffness $K_1$. Lower part, the inelastic neutron scattering data for LaB$_6$, from Ref. \onlinecite{Smith1985}, confronted with computed dispersions curves (full lines) in the limits of perfectly rigid (left) or soft (right) cages (see Appendix \ref{RelDisp}). The dashed lines on left represent the underlying flat local mode of the guest, at frequency $\omega_0$, and acoustic branch for empty cages with maximum frequency $\Omega_0$. }
\end{figure}

In case of strictly local modes of the guests inside their cages, a flat, low energy branch should appear when investigating the phonons dispersion. Actually, a large moving and coupled mass, as that of the rare-earth, cannot leave unaffected the cage and lattice. The consequences of its movement can be easily derived using a classical, harmonic model: a linear chain of springs and masses\cite{Christensen2008}. In the upper part of Fig. \ref{CageDisp}, a variant of linear chain model is detailed where, in addition to the springs linking the guest (mass $m$) to the cage (stiffness $k_0$), and the cage to its neighbors (stiffness $K_0$), a spring of stiffness $K_1$ links the two halves (masses $M/2$) of the cage. This allows to account for the elasticity of the cages in the description of a dispersion curve. The detail of the derivation of the dispersion relations is reported in Appendix \ref{RelDisp}. In the graphs of Fig. \ref{CageDisp}, this model is confronted with inelastic neutron scattering data showing the fourfold axis longitudinal modes of LaB$_6$\cite{Smith1985}. The experimental data is well described in the limit of infinitely rigid cages (left side, $K_1 \rightarrow +\infty$), but certainly not in the limit of soft cages (right side, $K_1=0$). Note that the only adjusted parameters are the two frequencies $\omega_0 = \sqrt{2 k_0 /m}$ and $\Omega_0 = \sqrt{4 K_0 /m}$, the mass ratio $m/M$ being fixed from the atomic masses of La and 99\% $^{11}$B enriched boron. It is thus shown that, for describing the two lowest dispersion curves of LaB$_6$, one can safely consider rigid cages. This should generalize to the hole series of the hexaborides as, for instance, the experimental data\cite{Kunii1997} for CeB$_6$ are very similar to those for LaB$_6$. These dispersion curves can be interpreted as the anticrossing between the low energy, flat branch of the guest vibration at frequency $\omega_0$ and the acoustic branch for a stiff lattice of empty cages with top frequency $\Omega_0$ (doted lines in Fig. \ref{CageDisp}). The gap that opens has a width directly related to the ratio $m/M$ between the guest and cage masses. For the light boron cages in the hexaborides, this ratio is large as is the gap\cite{Smith1985}, whereas for filled skutterudites, the much heavier cages result in a smaller separation \cite{Lee2006}. The experimental data for filled skuterrudites are scarce, by lack of large single crystals, but for those available (see curves in Ref. \onlinecite{Lee2006}), the lowest branches reproduce the pattern observed for the hexaborides: an anticrossing between a low energy "rattling" branch and the acoustic branch for a lattice of rigid cages.\newline
Using the harmonic approximation for an individual oscillator, the amplitude of the guest movement within the vibrational ground state is directly related to the levels separation $\Delta E = \hbar \omega_0$. Along a given axis, the low temperature r.m.s. deviation of the guest reads as: $\sigma = \hbar / \sqrt{2 m \Delta E}$. As shown in Fig. \ref{CageDisp}, a value for $\Delta E$ can be obtained by looking at the flattened end of the acoustic dispersion branch. In filled skuterrudites, this energy ranges between 4 and 9 meV \cite{Viennois2004, Lee2006, Iwasa2007}, while it is found slightly above 10 meV in rare-earth hexaborides ($\approx$ 13 meV in LaB$_6$ and CeB$_6$ \cite{Smith1985, Kunii1997}, $\approx$ 11 meV in PrB$_6$ \cite{Kohgi2006}). For these energies and an average lanthanide mass, the r.m.s. amplitude $\sigma$ ranges between 0.04 and 0.07 \AA. Unsurprisingly, the low temperature values are smaller than the room temperature estimates derived from the diffraction data.  Apart from this temperature effect, the spectroscopic and Debye-Waller investigations point to the same order of magnitude for the displacement inside the cages of skuterrudites and hexaborides, of about one tenth of {\AA}. \newline
This analysis shows that the rare-earth hexaborides and filled skuterrudites share specific traits that define them as magnetic cage materials: \newline
- large excursions of the guest inside the cage, with an amplitude of about 1/10 {\AA}, materializing in low frequency vibrations of the rare-earth.\newline
- a close crystallographic environment of the rare-earth, the cage, that can be viewed as essentially rigid when dealing with low energy phenomena.\newline

\subsection{The cage crystalline electric field}
The above analysis allows to consider the guest as mobile inside a rigid cage, with an excursion magnitude of a about one tenth of {\AA}. This is smaller than a rare-earth radius, which is in excess of one {\AA}, but of the same magnitude as a typical 4$f$ shell radius. In relative terms, the movement of the 4$f$ shell with respect to its cage environment is substantial: a priori, the CEF difference between a centered and offset position of the 4$f$ shell cannot be neglected.\newline

\subsubsection{Off-center crystal field}
In conventional compounds, the CEF is formalized by considering the point symmetry at the average position of the rare-earth, that here identifies with the cage center. The cage center has the symmetries of a cubic point group: $T_h$ in case of a filled skutterudite and $O_h$ for a rare-earth hexaboride. The expression for the center CEF hamiltonian, $\mathcal{H}_{CEF_0}$, describing the action of the CEF on the rare-earth ground state $J$ multiplet strictly reflects the point symmetry of the rare-earth site \cite{Takegahara2001, LLWolf1962}. In these high symmetries, only 4$^{th}$ and 6$^{th}$ order electric multipoles develop on the 4$f$ shell, but no quadrupoles: a degeneracy, larger than the Kramers' minimum, is preserved for some of the CEF levels (Fig. \ref{CagDiag}, left side). The rare-earth ion being mobile, out of the cage center the symmetry of its environment is drastically reduced (right side of Fig.~\ref{CagDiag}). The CEF acquires a dynamic character that, in the here considered rigid cage environment, is entirely due to the rare-earth movement. Considering the five orders of magnitude difference between the electron and rare-earth masses, it is the case to use a Born-Oppenheimer approximation, wherein the electronic 4$f$ wave functions continuously adapt to the position of the slowly moving rare-earth ion. Local CEF stationary 4$f$ states, associated with a corrected hamiltonian, are then a good approximation. This is accounted for by writing a position dependent, static, CEF hamiltonian, $\mathcal{H}_{CEF}(\bm{r})$, where, in addition to the main central term $\mathcal{H}_{CEF_0}$, a corrective off-center $\mathcal{H}_{CEF_d}$ is introduced:
\begin{equation}
\label{HCEF}
\mathcal{H}_{CEF}(\bm{r})=\mathcal{H}_{CEF_0}+\mathcal{H}_{CEF_d}(\bm{r})
\end{equation}

The continuous symmetry lowering, associated with the change in the position $\bm{r}$ of the rare-earth, is reminiscent of the treatment of magnetoelastic phenomena in rare-earth cubic compounds\cite{Mullen1974}: here, instead of being modulated by a strain, the corrective terms in the CEF hamiltonian have to reflect the excursion out of the cage center. In both cases, the main correction is the interference in the hamiltonian of 2$^{nd}$ order multipolar terms. This implies the emergence 4$f$ electric quadrupoles and the splitting of the non-Kramers CEF levels. The corrective $\mathcal{H}_{CEF_d}$ term have to develop with the distance $r$ to the cage center. Restricting to the lowest order, the correction is quadratic both in the coordinates of the ion and in those, relative, of the 4$f$ electrons. If the point group of the cage center is cubic, the crystal field correction, here written using the quadrupolar cubic irreducible representations \cite{MorinSchmitt1990}, has necessarily the form:\\
\begin{widetext}
\begin{equation}
\label{HJT}
\mathcal{H}_{CEF_d}(\bm{r})=-D^{\gamma} [(3 z^2-r^2) O_{2}^{0} + 3 (x^2-y^2) O_{2}^{2}]
-D^{\varepsilon} [(x \cdot y) P_{xy} + (y \cdot z) P_{yz} + (z \cdot x) P_{zx}]
\end{equation}   
\end{widetext}
where $x$, $y$ and $z$ are the components, along the cubic axes, of the displacement $\bm{r}$ of the rare-earth nucleus from the center of the cage. $\{O_{2}^{0}$, $O_{2}^{2}\}$ and $\{P_{xy}$, $P_{yz}$, $P_{zx}\}$ are the quadrupolar operators transforming, respectively, as the $\gamma$ ($\Gamma_3$) and $\varepsilon$ ($\Gamma_5$) cubic representations. In the $J$ manifold of the 4$f$ ion, they are conveniently written in terms of Stevens equivalents \cite{Stevens1952}. $D^{\gamma}$ and $D^{\varepsilon}$ are constants that, within a representation, define the magnitude of the coupling of the 4$f$ quadrupoles with the environment. In case of a displacement along a fourfold axis, only $D^{\gamma}$ is active, whereas along a threefold axis, it is $D^{\varepsilon}$.

\subsubsection{The broadening of the non-Kramers levels }
Inside the cage, as formalized by Eq. (\ref{HJT}), the CEF scheme is no longer a stable feature of the rare-earth, but depends on its position. The usual CEF scheme picture, with infinitely sharp energy levels, has to be abandoned: the non-Kramers cubic levels are broadened inside the cage, with an energy distribution that depends on the spatial distribution of the rare-earth. At low temperature, this distribution is characteristic of the cage oscillator ground state and, as the temperature is increased, thermally excited vibrations should further spread it. In systems with large displacement coupling constants $D^{\mu}$s and small CEF splitting at the cage center, this broadening of the CEF levels might be competitive with their separation.  This could explain anomalies reported in the neutron spectroscopy investigation of some filled skuttterudites, where CEF excitations are absent\cite{Galera2015}, even at low temperature, or vanish rapidly while increasing the temperature \cite{Iwasa2005, Park2008, Iwasa2009}.\newline
The thermal broadening of non-Kramers levels is certainly not exclusive to rare-earth cage compounds. However, in rare-earth systems with more common crystallographic structures, a substantial deviation from the high symmetry of the rare-earth site requires short wave distortions of its environment. This would involve high energy acoustic or optical phonons that are influential at temperatures typically competitive with the CEF levels spacing, above 100 K. At these temperatures, CEF effects are drastically reduced and there would be no point in considering the CEF scheme broadening. Reciprocally, at lower temperatures, the broadening is negligible and it is usually legitimate to consider the ideal symmetry of the rare-earth site.

\subsubsection{Effect of a non-Kramers ground state on the paramagnetic properties}
For temperatures lower than the maximum splitting of the ground state inside the cage, which defines a characteristic energy scale, the properties will reflect the uneven population of the local CEF states. As the temperature is reduced, the state with lowest energy of the split multiplet is favoured and the average magnetic entropy accordingly reduced, with consequences on the specific heat and the magnetic response (susceptibility).  In this temperature range, an analysis based on the central CEF scheme is inappropriate for describing the experimental value of the magnetic entropy, magnetic susceptibility and other CEF determined properties.

\subsubsection{The Jahn-Teller mechanism}
According to Eq. (\ref{HJT}), in case of a non-Kramers ground state, a splitting of this CEF multiplet develops quadratically with the distance $r$ to the center. This means that a lower energy electrostatic configuration can be achieved by moving away from the center. To be effective, this energy reduction also requires temperatures that reduce the statistical weight of the configurations with higher electrostatic energy: the energy scale associated with the cage splitting of the CEF ground state is also at play. This is the mechanism illustrated in Fig.~\ref{CagDiag}: as the temperature is decreased, a centrifugal kind of Jahn-Teller effect can be expected to develop in systems with a central non-Kramers CEF ground state.

\subsection{The cage potential}
It appears that the distribution of the rare-earth inside the cage is of critical influence on the CEF related properties, particularly in case of a non-Kramers, center ground state. The phonons dispersion analysis of Section \ref{Dispersion} shows that one can consider the rare-earth as exclusively coupled with a rigid cage, via the springs $k_0$ in Fig. \ref{CageDisp}. This means that the interaction between the cage and its guest can be treated separately, ignoring the rest of the crystal. In quantum mechanics, such an interaction is described by introducing a time independent cage potential for the guest. Solving the Schr\"{o}dinger equation for this potential well yields the sought after rare-earth distribution.

\subsubsection{Non-magnetic potential well}
The confinement of the rare-earth in the cage is not of magnetic origin. It is here formalized by introducing a potential well $V_0(\bm{r})$, where $\bm{r}$ refers to the displacement of the rare-earth nucleus from the cage center. The systems of interest are metals, where the cage framework is built from strongly bound atoms. Those, in contrast with the rare-earth ion, have small motion latitude, the cage being considered here as perfectly rigid (see discussion in Section \ref{Dispersion}).  As the rare-earth ion moves out of the center, its outer electrons reach those of the cage atoms: strongly repulsive forces develop, here simplified by considering an infinite barrier at the limits of the cage. This is not less realistic than a harmonic approximation and is of great practical interest when solving numerically the Schr\"{o}dinger equation (see Appendix \ref{NumSchroed}). The actual shape of the barrier shares, at minimum, the symmetry elements of the center point group and may be rather complex. As only a few, low energy vibration levels will be considered, a faithful description of the barrier is unnecessary. Instead, various degrees of approximation can be used, the most tractable ones being the cube and the sphere.\newline
Inside the cage, the bottom of the potential well cannot be flat, due to the electrostatic interaction between the rare-earth ion and charges from the neighboring atoms, bonds and conduction electrons. These competing contributions may result in a complex shape for the bottom potential. However, in the cage systems we consider, the maximum deviation of the guest from the center is about one tenth of a rare-earth ionic radius. This is not large with respect the the crystallographic distances and a lowest order description in the distance $r$ from the origin might be sufficient. Considering the central cubic point symmetry, this lowest order is necessarily isotropic, consisting in a simple quadratic term. To complement the infinite barrier, inside the cage the non-magnetic potential is described as: $V_0 (r)= \alpha \; r^2$, where $\alpha$ is a constant, a priori positive in a metal, accounting for the different electrostatic contributions.\newline
The $V_0 (\bm{r})$ term should have negligible temperature dependence in the temperature range of interest, below 100 K. In the following, it is considered as independent from the temperature.

\subsubsection{CEF contribution to the potential well}
\label{CEFcontrib}
Due to the degeneracy of a non-Kramers CEF ground-state, the 4$f$ electronic distribution can adjust to changes in its electrostatic environment. Here, the change is the consequence of the movement of the ion inside the cage and will contribute to the cage potential well with a specific 4$f$ term. This is formalized by the hamiltonian term of Eq. (\ref{HJT}), that has to be translated into an extra mean-field potential term $V_{4f}(\bm{r}, T)$. In the original Born-Oppenheimer approximation, this question is treated adiabatically, considering only the lowest electronic energy level as the potential term. In the cage context, this lowest energy is not well separated from the excited electronic levels: due to the degenerate center CEF ground state, the energy separation goes to zero at the origin, thus realizing a conical intersection. One has then to face the difficulties of a non-adiabatic approach. One simplification comes from the fact that the rare-earth ion and its cage are not an isolated molecule, but belong to a crystal, moreover a metallic one. Then, even if the movement of the rare-earth is so slow that the dynamic mixing of the local CEF states can be neglected, one has to consider the perturbing effect of the environment. In particular, frequent collisions should occur with conduction electrons, inducing transitions between the local CEF states. As a result, the electrostatic forces exerted on the rare-earth ion and associated potential term will rapidly fluctuate. The fast electronic fluctuations will have no effect on the massive rare-earth which will be sensitive to an average that represents the $V_{4f}(\bm{r}, T)$ mean-field. For an extremely slow rare-earth movement, the statistics of the local 4$f$ states will approach a Boltzmann distribution defined by the temperature of the crystal. At constant temperature $T$, the additional work required to move the rare-earth ion, because of its evolutive 4$f$ aspherical distribution, equals the variation of free energy associated with the 4$f$ shell. In this non-adiabatic approximation, at a given temperature $T$, the magnetic part of the potential identifies with the local 4$f$ free energy:
\begin{equation}
\label{V4f}
V_{4f}(\bm{r}, T)= -k_{\text{B}}\;T\; ln \frac{Z(\bm{r},T)}{2J+1}
\end{equation}
where $Z(\bm{r},T)$ is a local partition function for the 4$f$ electronic states at a point $\bm{r}$ and temperature $T$. The division by $2J+1$ is required in order to have a zero $V_{4f}(\bm{r}, T)$ at $r=0$. In contrast with $V_{0}$, the $V_{4f}$ potential is clearly temperature dependent. Indeed, at temperatures high with respect to the splitting inside the cage, all local CEF are equally populated and the $V_{4f}$ potential term flattens inside the cage. Conversely, at low temperature, $V_{4f}(\bm{r})$ will follow the energy dependence of the lowest CEF level, decreasing quadratically with respect to the components of $\bm{r}$.

\section{Spherical proof of concept}
In principle, the discussion in the previous section allows to define the total cage potential at a given temperature:
\begin{equation}
\label{v(r)}
V(\bm{r}, T)=V_0(\bm{r})+V_{4f}(\bm{r}, T)
\end{equation}

\begin{figure}
\includegraphics[width=\columnwidth]{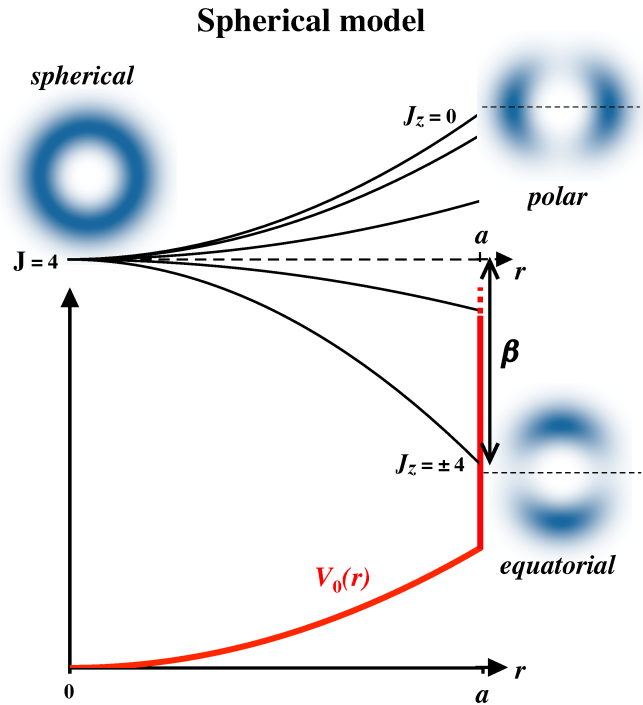}
\caption{\label{SpheMod} Spherical model for a rare-earth ion with excursion latitude $a$ inside a rigid cage. The lower part shows the non-magnetic potential $V_0({r})$ consisting in a quadratic bottom and infinite barrier at $r=a$. The upper part shows the splitting of the rare-earth multiplet (here exemplified by the Pr$^{3+}$ ion, $J = 4$), that yields a magnetic potential $V_{4f}(r, T)$, at a distance $r$ from the center along the quantization $z$ axis. Here, the lowest energy is achieved for the doublet $J_z = \pm 4$ with "equatorial" distribution of the 4$f$ electrons and $\beta$ decrease in the electrostatic energy at $r = a$. The energy scales for the splitting and $V_0({r})$ are independent.}
\end{figure}
From there, describing the movement of the enclosed rare-earth ion requires to solve the time independent Schr\"{o}dinger equation for the mass $m$ of the rare-earth ion inside the potential well described by $V(\bm{r}, T)$. To describe the properties of a specific rare-earth system, one still has to detail the $V_0(\bm{r})$ term, which requires to choose an approximant for the cage shape, defining the infinite barrier in $V_0(\bm{r})$, and select a value for the constant $\alpha$. As regards the magnetic part of the potential well, the knowledge of the cage center CEF scheme is required or, at least, for a low temperature description, an identification of the CEF ground state. Then, one has to select values for the $D^{\gamma}$ and $D^{\varepsilon}$ displacement-quadrupole couplings constants, involved in Eq. (\ref{HJT}). After that, there remains the technical difficulty of integrating the Schr\"{o}dinger equation for the possibly complex potential well.\newline

Here, our purpose is not to investigate a specific compound, but to identify the physical consequences of a position dependent CEF in the cage. This can be achieved using a further simplified model, provided it retains these essential features: \newline
- the magnetic ion is trapped inside a cage, with high point symmetry at the center. \newline
- the electronic ground state at the cage center is degenerate. \newline
- the degenerate electronic states are split as the ion moves out of the center. \newline
By reducing the physical system to a spherical cage enclosing a rare-earth ion, the first condition is met. Moreover, as the inner electric field produced by a uniformly charged sphere cancels, the 4$f$ electronic ground state should retain the full $2J+1$ multiplet degeneracy. In order to lift the orbital degeneracy out of the cage center, one has to consider an additional, spherical charge density inside the cage. In a metal, this charge would correspond to that of the conduction electrons. In order to introduce temperature dependent effects, the cage and its guest also need to be coupled with a thermostat at $T$. This is also a role that can be deferred to the conduction electrons. \newline

\subsection{Spherical potential well}
\label{SpherWell}
The spherical simplification reduces the non-magnetic $V_0(\bm{r})$ potential to a form consisting in (see Fig. \ref{SpheMod}, lower part): \newline
- an infinite spherical barrier at radius $a$, \newline
- a restoring force, restricted to a quadratic term in the potential for $r < a$: $V_0 (\bm{r})= \alpha \; r^2$, where $\alpha$ is positive. \newline
The radius $a$ of the cage and the mass $m$ of the guest define the practical units for:\newline
- length, $a$. Instead of using $\bm{r}$, the position in the cage is below referred to by $\bm{\rho}=\bm{r}/a$, in $a$ unit.\newline
- energy, $e.u. =\frac{\hbar^2}{2m a^2}$, by reference to the energy levels of an infinite spherical well with radius $a$.\newline
The $V_{4f}(\bm{r}, T)$ term reflects the electrostatic interactions between the 4$f$ shell and the conduction electrons. In the present spherical simplification, it should not reflect the quadrupolar anisotropy of Eq. (\ref{HJT}), that arises in cubic symmetry for independent $D^{\gamma}$ and $D^{\varepsilon}$ coupling constants. Taking the quantization $z$ axis along the displacement direction, the isotropic reduction of the hamiltonian term of Eq. (\ref{HJT}) reads as :
\begin{equation}
\label{HJTr}
\mathcal{H}_{CEF_d}(\rho)=-  \mathcal{D} \; \rho^2 \;O_{2}^{0}=- \mathcal{D} \; \rho^2\;[3J_z^2-J(J+1)] 
\end{equation}
where $ \mathcal{D} = 2\;D^{\gamma} a^2= 24\;D^{\varepsilon} a^2$.

This hamiltonian describes the splitting of the $J$ multiplet outside the cage center (see Fig. \ref{SpheMod}, upper part). The eigenstates coincide with the $|J, J_z \rangle$ states, the ones with opposite $J_z$ projections being degenerate. Depending on the sign of $\mathcal{D}$, the local CEF ground state at $\rho$ will either correspond to the maximum projection doublet, $J_z = \pm J$ or to the minimum, $J_z = 0$ singlet or $J_z =\pm \frac{1}{2}$ doublet. Following Stevens' equivalent operators method\cite{Stevens1952}, the 4$f$ electrons quadrupolar component along the displacement axis, $Q_{zz} =  \left\langle 3z^2 - r^2 \right\rangle$, is directly related to the $J_z$ projection:
\begin{equation}
\label{Stevens}
Q_{zz} = {\alpha_J} \left\langle{R_{4f}}^2\right\rangle \left\langle {3{J_z}^2 - J(J + 1)} \right\rangle 
\end{equation}
, where $\alpha _J$ is the Stevens' second-order constant and $\left\langle {R_{4f}}^2 \right\rangle$ the second moment for the 4$f$ radial wave function of the considered rare-earth. From Eq. (\ref{HJTr}), it appears that the energy extrema correspond to an equatorial, or polar distribution of the 4$f$ electrons along the displacement axis. To identify the configuration of lowest energy, more detail about the conduction electrons density is required. If a net excess of negative charge lies well inside the average 4$f$ radius, reduced to a central negative charge, a simple electrostatic calculation shows that an equatorial distribution of the 4$f$ electrons is favored. This means that the coupling constant $\mathcal{D}$ has to be negative for $\alpha _J$ positive (which occurs only for the heaviest tripositive rare-earth) and, respectively positive, for $\alpha _J$ negative (which is the case for most tripositive rare-earth ions). In the following, it is assumed that the equatorial configuration is of the lowest energy, as represented on Fig. \ref{SpheMod}. \newline
For a given value of $\mathcal{D}$, $T$ and $\rho$, the local partition function $Z(\rho,T)$ can be computed, then the potential term, according to Eq. (\ref{V4f}). Once the spherical potential $V(\rho, T)$ is defined, one can turn to solving the time independent Schr\"{o}dinger equation for the enclosed nucleus at a given temperature $T$. The angular part of the eigenfunctions are the spherical harmonics $Y_{l}^{m}(\theta,\varphi)$, whereas the radial part $R_{n,l}(\rho)$ and energies $E_{n,l}$ require solving a differential equation specific to the considered potential. Except for very specific $l$ values and potential shapes, one is forced to resort to a numerical treatment (see Appendix \ref{NumSchroed}).\newline
In the following, the solutions to the Schr\"{o}dinger equation are labeled using the usual atomic notations $n[l]$, the sorted sequence of the levels being: 1$s$, 2$p$, 2$d$ etc.. In all physically relevant cases this energy sequence is maintained, with a 1$s$ vibration ground-state. 

\subsection{Thermodynamic and magnetic consequences} 
\label{SecThermo}
\subsubsection{Energy scales} 
In the hypothesis of a conduction electrons density favouring an equatorial distribution of the 4$f$ electrons around the displacement axis, the lowest energy corresponds to the $O_{2}^{0}$ value $Q_0 =2 J^2 -J$, for $\alpha _J$ negative (doublet, $J_z =\pm J$). For $\alpha _J$ positive, $Q_0 =-J(J+1)$ for an integer $J$ (singlet, $J_z =0$), or $Q_0 = 3/4-J(J+1)$ for a half-integer $J$ (doublet, $J_z =\pm 1/2$). One can introduce the constant:
\begin{equation}
\label{beta}
\beta = \mathcal{D} Q_0 
\end{equation} 
that equals the decrease in energy of the lowest 4$f$ level when the rare-earth has maximum deviation from the center (see Fig. \ref{SpheMod}). As the temperature approaches the range of $\beta$, the local probabilities of occupation will start to segregate between the 4$f$ levels. This has necessarily an impact on the thermodynamic and magnetic properties which acquire a specific temperature dependence for $T$ lower than $\beta$. Estimating the order of magnitude of $\beta$ is crucial in order to decide wether it interferes with the energy scale of the vibrations. To this regard, the only quantitative experimental indication comes from the example of CeB$_6$, in which an increase of 1.24 meV in the energy of the $\Gamma_8$ - $\Gamma_7$ Raman CEF excitation \cite{Zirngiebl1984} is observed when cooling down from 300 K to 4 K. Ascribing this to the average lowering of the cage split $\Gamma_8$ level (the $\Gamma_7$ doublet cannot be split), the magnitude of the $\beta$ equivalent in the CeB$_6$ case should be about 2 meV. As regards the energy scale of the vibration, for rare-earth hexaborides, experiments show that the $\Delta E$ separation of the lowest vibration levels is slightly above 13 meV for LaB$_6$ (see Fig. \ref{CageDisp}). This CeB$_6$ example thus yields a 15 \% estimate for the ratio between $\beta$ and $\Delta E$. In this case, the effect of the splitting of the central ground state multiplet will manifest itself at temperatures much lower than the $1s-2p$ separation of the spherical levels. This allows a further simplification, wherein the only considered cage distribution is that of the lowest 1$s$ level. \newline
As regards the calculation of macroscopic observables such as the internal energy, entropy, magnetic susceptibility etc., note that the hypothesis of a local Boltzmann distribution (introduced in section \ref{CEFcontrib} for computing the magnetic contribution to the cage potential well) is not required. The measured systems contains an extremely large number of equivalent cages which, considered as independent, realize a large ensemble for which a Boltzmann distribution is certainly applicable. As shown below in Section \ref{Centrifug}, in relative terms, the magnetic effect on the potential is small.

\subsubsection{The praseodymium example} 
At a site of cubic symmetry, the CEF ground state of an ion with integer value of $J$ can be 1, 2 or 3-fold degenerate\cite{LLWolf1962}. In case of a half-integer $J$, the ground state can be 2 or 4-fold degenerate. The here relevant cubic CEF ground states are non-Kramers, i.e. carry an excess of orbital degeneracy. In case of an integer values of $J,$ they are thus 2 or 3-fold degenerate and 4-fold degenerate for a half-integer $J$. If, instead of a pseudo-spin, one considers a real rare-earth ion inside the spherical cage, the central $2J + 1$ degeneracy is necessarily higher than that of a cubic, non-Kramers CEF ground state. Nevertheless, for the sake of a quantitative illustration, we arbitrarily select the example of a $J = 4$ multiplet that correspond to the ground multiplet of a Pr$^{3+}$ ion (9-fold degenerate at the center). As the second order Stevens coefficient $\alpha _J$ is negative for Pr$^{3+}$, an equatorial distribution of the 4$f$ electrons corresponds to a positive $\left\langle {O_{2}^{0}} \right\rangle$. This means that the local lowest energy level is the $J_z = \pm 4$ doublet and that the zero temperature limit of $\left\langle {O_{2}^{0}} \right\rangle$ is $Q_0 = 28$ (see Fig. \ref{SpheMod}). Moreover, to get the minimum energy for this 4$f$ distribution, the $\mathcal{D}$ constant in Eq. (\ref{HJTr}) has to be positive. Then, one has to choose a value for $\mathcal{D}$. For the most simple, flat bottom, spherical well, the 1$s$-2$p$ separation is 10.33 $e.u.$, which means that a simple value $\beta = 2 \; e.u.$ agrees with the order of magnitude derived from the CeB$_6$ example. In order to have $\beta = 2\; e.u.$ for a Pr$^{3+}$ guest, one has $\mathcal{D} =\beta / Q_0 = 1/14\; e.u.$.

\subsubsection{Calculation and results} 
\begin{figure}
\includegraphics[width=\columnwidth]{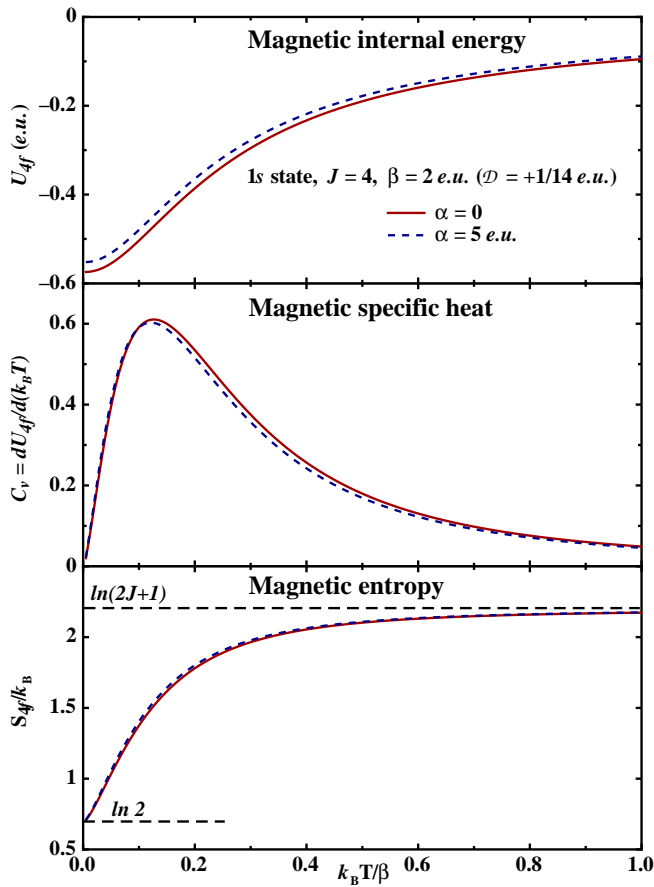}
\caption{\label{Thermo} Computed thermodynamic consequences, as function of the reduced temperature $k_{\text{B}} T/\beta$, for a cage split $J = 4$ multiplet in the spherical 1$s$ vibrational ground state. The full line curves are obtained for a non-magnetic potential with flat bottom ($\alpha = 0$) whereas the dashed lines are for a quadratic bottom ($\alpha = 5 \;e.u.$). $\beta$ is the amplitude of the energy decrease for the lowest CEF level at the cage limit. The upper part gives the temperature variation of the 4$f$ shell contribution to the internal energy. The middle and lower frames respectively show the associated specific heat and magnetic entropy. }
\end{figure}

Another required parameter for the calculation is $\alpha$, that defines the quadratic bottom for the non-magnetic potential $V_0(r)$. There is no experimental data that could help estimate the magnitude of $\alpha$. Two values, 0 for a flat bottom and 5 $e.u.$ for a pronounced harmonic bottom, relatively to the 1$s$-2$p$ separation, are used in the following calculations. \newline
Once the potential is defined at a given temperature $T$, solving the radial Schr\"{o}dinger equation (see Appendix \ref{NumSchroed}) yields the radial wave function $R_{1,0}(\rho)$. The 4$f$ partition function at $\rho$, allows to define the local values for the internal energy, $U_{4f}(T, \rho)$, entropy, $S_{4f}(T, \rho)$, and specific heat, $C_{V_{4f}}(T, \rho)$. The averaged values, $U_{4f}(T)$, $S_{4f}(T,)$ and $C_{V_{4f}}(T, \rho),$ for the 1$s$ state are then obtained by numerical integration over the radial distribution. For instance, in the case of $U_{4f}(T)$:
\begin{equation}
\label{average}
U_{4f}(T) = \int_{0}^{1} \; 4 \pi \rho^2\; |R_{1,0}(\rho)|^2 \; U_{4f}(T, \rho) \; d\rho
\end{equation}
Fig. \ref{Thermo} gives the temperature dependencies of these quantities, in the range defined by the $\beta$ value. The plot of the constant volume specific heat $C_V$ is obtained from a numerical derivative of $U_{4f}$. The chosen value for $\beta$ has little impact on the potential well, then on the temperature variation of the 1$s$ vibration energy: in this temperature range, the variation of the total internal energy is essentially that of $U_{4f}$.

In the upper part of Fig. \ref{Thermo} one can observe how the average 4$f$ energy is lowered for temperatures below $\beta /k_{\text{B}}$. In a real system, with a non-Kramers CEF ground state at the center, this decrease in energy of the CEF ground state results in a larger transition energy to an un-split excited CEF level. This is precisely what is observed via neutron and Raman scattering experiments in the example of CeB$_6$\cite{Zirngiebl1984}. Note that the shift in energy in this spherical illustration represents only 27 \% of the $\beta$ value and about 5 \% of the energy difference between the lowest 1$s$ and 2$p$ vibration energies. A better agreement with the order of magnitude derived from CeB$_6$ example would require a larger $\beta$ value of about 6 $e.u.$.  As regards the specific heat, the 4$f$ energy lowering results in a Schottky-like anomaly that peaks at a temperature between 0.1 and 0.2 $\beta /k_{\text{B}}$. In systems where an ordering occurs at temperatures competitive with $\beta /k_{\text{B}}$, only the higher temperature part of the peak will appear on the specific heat curves, the peak itself being concealed by the order. This is typically the case in rare-earth hexaborides where most elements order antiferromagnetically.\newline
Low temperature Schottky-like anomalies are reported for rare-earth filled skutterudites, notably for praseodymium compounds with non-magnetic CEF ground state as PrOs$_4$Sb$_{12}$\cite{Maple2002} or PrOs$_4$P$_{12}$ \cite{Matsuhira2005}. These anomalies are interpreted as the effect of a low lying CEF level above the $\Gamma_1$ singlet ground state. As illustrated here, an alternative would be to consider an isolated, cage split, non-magnetic $\Gamma_{23}$ doublet ground state. Schottky anomalies are also observed to survive the ferromagnetic order in neodymium compounds such as NdOs$_4$Sb$_{12}$ \cite{Ho2005}, NdOs$_4$As$_{12}$ \cite{Cichorek2014} or NdRu$_4$As$_{12}$\cite{Rudenko2016}. In all these examples, the crystal field ground state seems to be a non-Kramers quadruplet, therefore likely to be split in the cage context.\newline
The associated entropy (Fig. \ref{Thermo}, lower part), shows that a temperature level of $\beta /k_{\text{B}}$ is required to recover the entropy of the $J=4$ multiplet. A prolonged increase of the entropy in the paramagnetic range, with difficulties for achieving the well defined degeneracy of a CEF ground state, is a frequent trait of rare-earth cage compounds. In the illustrating case of the $J=4$ multiplet, the choice of a positive value for $\mathcal{D}$ defines the doublet $J_z = \pm 4$ as the local CEF ground state at any point in the cage, except the center. This results in an average magnetic entropy that doesn't go to zero but to the finite value $k_{\text{B}}\,ln\,2$ at zero temperature. Conversely, a negative $\mathcal{D}$ would have resulted in a singlet local CEF ground state and vanishing magnetic entropy at zero Kelvin.\newline
Another information brought by the calculation is the little influence of the bottom of the non-magnetic cage potential, expressed via the values 0 (full line in Fig. \ref{Thermo}) and 5 $e.u.$ (dashed line) for $\alpha$. The positive $\alpha$ slightly reduces the extension of the 1$s$ wave functions, diminishing, at a given temperature, the CEF energy gain. However, the amplitude of the effect and overall aspect of the curves are very similar. The essential parameters of the model are the size of the cage $a$ and the CEF splitting amplitude $\beta$. Both define separate energy scales, $a$ acting on the guest vibration energy $\Delta E$ whereas $\beta$ determines the lower energy CEF effects.

\subsubsection{Magnetic susceptibility} 
\begin{figure}
\includegraphics[width=\columnwidth]{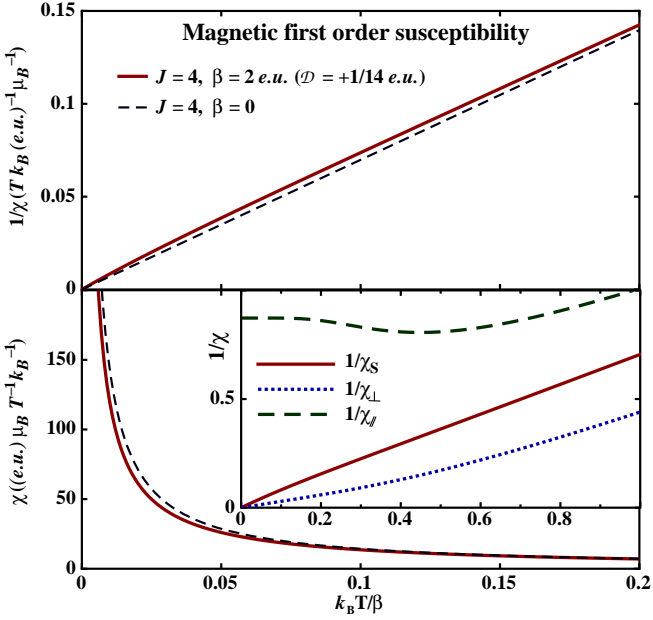}
\caption{\label{Mag} Effect of the cage split $J = 4$ multiplet on the magnetic susceptibility $\chi$, as function of the reduced temperature $k_{\text{B}} T/\beta$. The upper graph shows the inverse magnetic susceptibility, the lower one, the direct susceptibility. The full line gives the result of the calculation for the 1$s$ state with $\mathcal{D} = 1/14\; e.u.$ and $\alpha = 0$ (there is no visible difference for $\alpha = 5\;e.u.$ ), while the dashed line gives the Curie law for $J = 4$. The inset shows the inverse susceptibilities, parallel, $\chi_{\|}$, and perpendicular, $\chi_{\bot}$, to the displacement axis, as well as the resulting spherical average $\chi_S$, used as a standard curve (see text).}
\end{figure}

One most visible influence of the CEF scheme, when investigating rare-earth compounds, is the change in the magnetic susceptibility with respect to the degenerate $2J+1$ ground state multiplet. Similarly, an effect on the susceptibility can be expected in the $\beta /k_{\text{B}}$ temperature range, where the population of the 4$f$ states evolves. To compute the magnetic susceptibility in the cage context, a Zeeman term needs to be added to the local CEF hamiltonian of Eq. (\ref{HCEF}). In the spherical simplification, the CEF term at radius $\rho$ reduces to the uniaxial $O_2^0$ term (Eq. \ref{HJTr}) and has to be complemented with the Zeeman perturbation in order to write the 4$f$ hamiltonian at $\rho$:
\begin{equation}
\mathcal{H}(\rho)=- \mathcal{D} \; \rho^2\;\frac{\hbar^2}{2m a^2} O_2^0 +\mu _0\;\mu_{\text{B}}\;{g_J}\;\bm{H} \cdot \bm{J}
\end{equation}
where $g_J$ is the Land\'e factor and $\bm{H}$ the applied magnetic field.
In the uniaxial symmetry, the local susceptibility tensor reduces to the susceptibilities parallel $\chi_{\|}$ and perpendicular $\chi_{\bot}$ to the displacement axis. At given temperature $T$ and position $\rho$,  $\chi_{\|}(T , \rho)$ and $\chi_{\bot}(T , \rho)$ can be computed from a perturbation approach or, as done here, via a linearization of the magnetization curve obtained from numerical diagonalization and Boltzmann statistics. Considering only the 1$s$ vibration state, the effective susceptibility results from a spherical average over all displacement directions. The averaged contribution from a radius $\rho$ reads as:
\begin{equation}
\chi(T, \rho)= \frac{\chi_{\|}(T , \rho)+2\chi_{\bot}(T , \rho)}{3}
\end{equation}
The total, isotropic, susceptibility $\chi(T)$ is then obtained by a second average of $ \chi(T, \rho)$ over the 1$s$ radial distribution at temperature $T$ (see the example of Eq. (\ref{average})).
In practice, it is not necessary to compute $\chi(T, \rho)$ for each radius value. Since the scale of the local CEF scheme is entirely defined by $\mathcal{D} \rho^2$, a standard curve $\chi_{S}(T)$ can be computed once for all and then adapted to any particular value of $\rho$. In the present case, the calculation is done for $\mathcal{D} = 1/14\; e.u.$ and $\rho=1$ over and extended temperature range (inset of Fig. \ref{Mag}), thus defining $\chi_{S}(T)$. From this standard curve, $\chi(T, \rho)$ for $\rho< 1$ is scaled as:
\begin{equation}
\chi(T, \rho)=\frac{\chi_{S}( T / \rho^2)}{\rho^2} 
\end{equation}

The results of Fig. \ref{Mag} show a moderate effect of the cage splitting, with respect to a reference Curie law for $J = 4$. The difference is visible only for temperatures much lower than $\beta /k_{\text{B}}$. As usual for a CEF effect in a high symmetry system, a reduction of the susceptibility is observed. Due to the hypothesis of a positive $\mathcal{D}$ value, the split multiplet has a local $J_z = \pm 4$, magnetic doublet ground state, that maintains a large susceptibility.  A very different picture would arise for a negative $\mathcal{D}$ value, that selects a non-magnetic local ground state. The effect of the quadratic bottom of the potential well is here imperceptible, at least in the considered cases $\alpha = 0$ and $\alpha = 5\;e.u.$ .

\subsection{The centrifugal Jahn-Teller effect}
\label{Centrifug}

\subsubsection{Zero temperature limit} 
In order to identify the effects of the $V_{4f}$ potential term, one can focus on the zero temperature limit, where it has maximal amplitude and simplest analytical form. For $T$ going to zero, in the set of quadrupolar values at $\rho$, $Q_0$ with its lowest energy is statistically dominant. In the zero temperature limit, the magnetic potential simplifies to:  $V_{4f}(\rho)=-\beta \rho^2$.
Then, at $T=0$ and for $\rho<1$, the total cage potential writes as:
\begin{equation}
\label{TotPot}
V(\rho, 0)=(\alpha-\beta )\;\rho^2=W\; \rho^2 
\end{equation}

\begin{figure}
\includegraphics[width=\columnwidth]{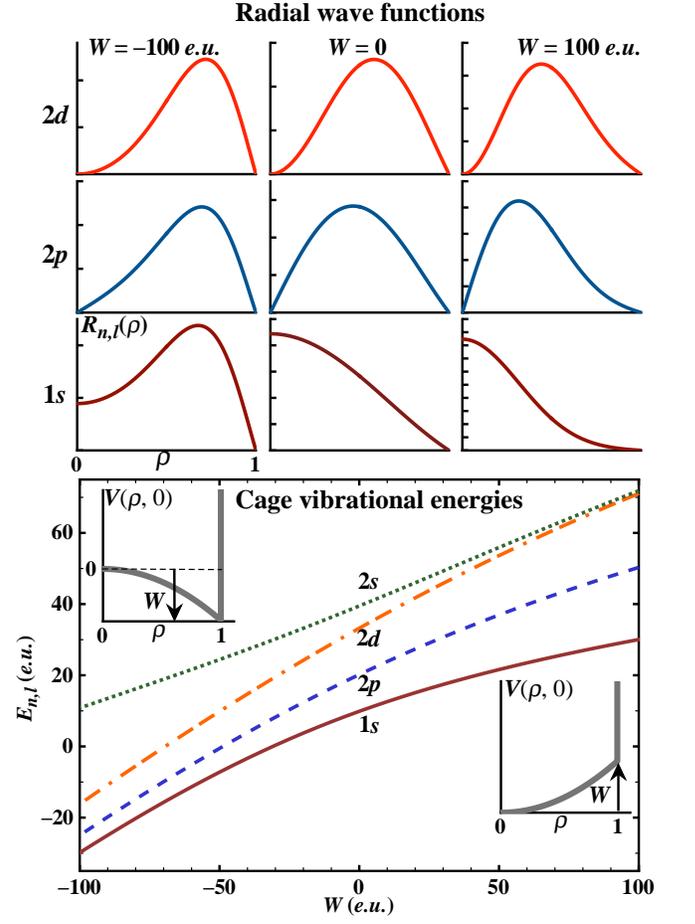}
\caption{\label{VarW} Zero temperature effect of the spherical well quadratic bottom, defined by the constant $W$. Lower part: Energies of the 4 lowest oscillator levels, $1s$, $2p$, $2d$ and $2s$, as functions of $W$ with the insets representing, as functions of the reduced radius $\rho$, wells for positive and negative $W$. The CEF contribution results in a $\beta$ shift of $W$ towards negative values. Upper part: sketches of the radial wave functions for the $1s$, $2p$ and $2d$ states, for a flat bottom ($W=0$) and for opposite strong convexities ($W=100$ and $W=-100$ $e.u.$).}
\end{figure}

Fig. \ref{VarW} shows the dependence on $W$ of the four lowest energy levels: 1$s$, 2$p$, 2$d$ and 2$s$. For large positive $W$ values, the system tends to harmonicity, with a constant level separation and "accidental" 2$d$-2$s$ degeneracy. The centrifugal 4$f$ term shifts $W$ the opposite way, towards negative values by the amount of $\beta$, which results in:\newline
- a reduced energy difference between the three lowest levels, particularly between the singlet 1$s$ ground state and the first excited triplet 2$p$ states.\newline
- radial wave functions that adapt to the centrifugal term. The 1$s$ ground is the most affected (see the upper part of Fig. \ref{VarW}) as, from Eq. \ref{SchroeRad}, the centrifugal effect is exclusively of magnetic origin in case $l=0$.\newline
The concomitant reduction of energy separation and increase in the wave functions overlap increases the multipolar susceptibilities of the rare-earth distribution. In particular, the dipolar and quadrupolar susceptibilities, respectively involving $1s$ to $2p$ and $1s$ to $2d$ matrix elements. This should facilitate low temperature phenomena that involve a redistribution of the rare-earth ion inside the cage. For instance, the dipolar susceptibility is at play in the antiferromagnetic range of rare-earth hexaborides, where exchange induced displacements of the rare-earth (i.e. dipoles) determine the magnetic wave-vector and the first-order kind of the magnetic transition\cite{Amara2005, Amara2010}. The cage quadrupoles, analogously to the 4$f$ ones\cite{Mullen1974}, should couple with the crystal strain modes and, via their associated susceptibilities, influence the elastic properties. This coupling with the rare-earth cage distribution also applies for a volume strain. Despite the above consideration of a perfectly rigid cage, in a real elastic system, an evolution in the guest radial distribution (see the sketches for the 1$s$ radial wave function in Fig. \ref{VarW}) will impact the cage and lattice volumes. There is scarce experimental data as regards the paramagnetic thermal expansion of filled skutterudites, but an investigation of rare-earth hexaborides shows noticeable differences between the non-magnetic LaB$_6$ and other elements in the series\cite{Sirota2000}. This changes are observed below T = 50 K, in a range where CEF effects should be absent, at least for CeB$_6$, PrB$_6$ and NdB$_6$, due to their large CEF level spacing \cite{Zirngiebl1984, Loewenhaupt1986}.\newline
The centrifugal term should also affect the lowest acoustic and optical phonon branches (see Fig. \ref{CageDisp}): the reduced 1$s$-2$p$ spacing at $T=0$ means that, for all phonons involving the vibration of the rare-earth inside its cage, some softening will develop as the system is cooled. A possibly related effect is observed in PrOs$_4$Sb$_{12}$\cite{IWASA2006} and in other filled skutterudites \cite{Iwasa2007}.

\subsubsection{Thermal dependence}
\begin{figure}
\includegraphics[width=\columnwidth]{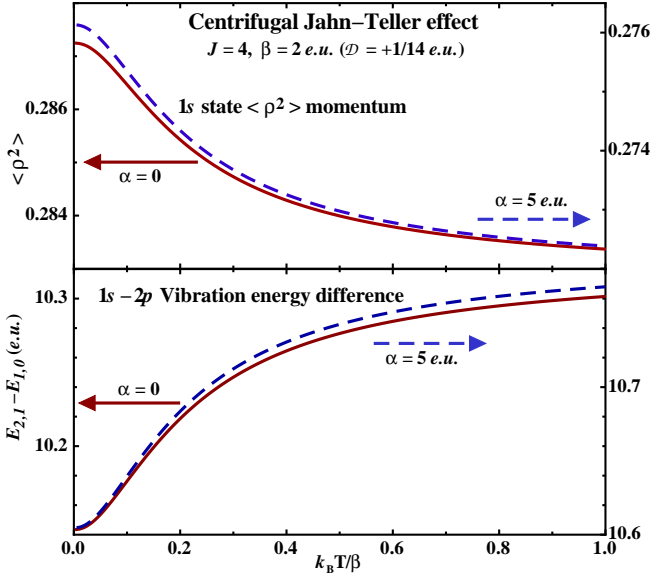}
\caption{\label{VarT} Illustrations of the centrifugal Jahn-Teller effect in the spherical simplification within the 1$s$ state, for $J = 4$, $\beta = 2\;e.u.$ and two values, 0 (full lines) and 5 $e.u.$ (dashed lines), of $\alpha$. The upper part materializes the centrifugal effect via the temperature dependence of the second order radial moment $\langle \rho^2 \rangle$ for the cage guest. The lower part shows the softening effect on the energy difference between the lowest vibration levels 1$s$ and 2$p$. The left vertical scales are for the flat bottom case, $\alpha = 0$, whereas the right ones are for $\alpha = 5 \;e.u.$. Left and right scales do not differ in amplitude, but are adapted by a shift.}
\end{figure}

To identify the above effects, the amplitude of the centrifugal term has been deliberately exaggerated: the horizontal scales of Fig. \ref{VarW} go well beyond the orders of magnitude considered in Section \ref{SecThermo}. Here, the same $J=4$, $\beta = 2\;e.u.$ and $\alpha = 0\; or \;\alpha = 5\;e.u.$ values are used to provide more quantitative information as regards the centrifugal effect and its temperature dependence. This relies on the more delicate, non-adiabatic, approximation that allows to individually define a potential for the considered cage (Eq. (\ref{V4f})).\newline
Fig. \ref{VarT} shows, as a direct illustration of the centrifugal effect, the temperature dependence of the second order radial moment $\langle \rho^2 \rangle$ for the 1$s$ state. The radial distribution expands as the temperature goes below $\beta /k_{\text{B}}$. For $\beta = 2\;e.u.$, the amplitude of the relative change in $\langle \rho^2 \rangle$ reaches about 1.5 \%. This amplitude is similar for the two considered non-magnetic bottoms, defined by $\alpha = 0$ and $\alpha = 5\;e.u.$. The absolute values of $\langle \rho^2 \rangle$ for $\alpha = 5\;e.u.$ are logically smaller, because of the associated restoring force that opposes the Jahn-Teller centrifugal effect.\newline
This seems too small to be investigated by diffraction techniques through a Debye-Waller analysis, but high precision dilatometry might be an option. Indeed, as the cage and crystal cannot be infinitely rigid, an isotropic change in the rare-earth distribution has necessarily an impact on the crystal volume. In an elastic cubic lattice of cages, since a volume strain $\varepsilon^{\alpha} = \varepsilon_{xx}+ \varepsilon_{yy}+\varepsilon_{zz}$ and the second order radial moment of the guest $\langle \rho^2 \rangle =\langle x^2 \rangle+ \langle y^2 \rangle+ \langle z^2 \rangle$ transform identically, they can be phenomenologically related\cite{Callen1963}. At the lowest order description, a small change in $\langle \rho^2 \rangle$ is linearly related to a volume strain: $\varepsilon^{\alpha}= \kappa (\langle \rho^2 \rangle-\langle \rho^2_0 \rangle)$, where $\langle \rho^2_0 \rangle$ is the second order radial moment in absence of the centrifugal effect. At this point, one cannot guess the magnitude and sign of the constant $\kappa$. Nevertheless, for a sufficiently large $\kappa$, a specific volume change and associated thermal expansion anomaly should be detected in a cage system with a non-Kramers CEF ground state.\newline
The lower part of Fig. \ref{VarT} shows another effect of the centrifugal term in the cage potential, that is the decrease of the energy differences between the lowest lying vibrational states. The graph displays the most relevant difference between the 1$s$ ground state and 2$p$ first excited level. This difference defines the vibration frequency of the "rattler" that, at low temperature, can be identified with $\omega_0$ in the harmonic approximation of Section \ref{Dispersion}. Here, for $\beta = 2\;e.u.$, a softening of about 1.5 \% is predicted, that should directly reflect on the low temperature dispersion curves (see Fig. \ref{CageDisp}), particularly on the flattened part of the "acoustic" branch and, at $q=0$, on the "optical" branch.
Resolving a 1.5 \% softening is difficult using inelastic neutron scattering and infrared spectroscopy might be better adapted for detecting the small energy shift in the optical branch at $q=0$.

\section{Summary}
Inside a high symmetry cage, an offset position of a rare-earth ion splits its orbitally degenerate electronic levels. These levels thus acquire a dynamical width that reflects the distribution of the guest inside the cage. In case of an orbitally degenerate CEF ground state at the cage center, the associated energy width defines a temperature range, in the paramagnetic state, where very specific phenomena take place. Within this range, as the temperature is lowered, the lowest level of the split multiplet becomes statistically dominant. This allows a reduction of the 4$f$ energy. As this energy gain is obtained for an offset position, a centrifugal Jahn-Teller force is simultaneously exerted on the guest.\newline
In order to investigate the physical consequences of a cage split electronic ground-state, a simplified, spherical model has been used, where a $J = 4$ multiplet enacts the degenerate level at the cage center. The experimental data on CeB$_6$ offers a quantitative frame for this description, fixing a value for the energy width of the multiplet, that is much smaller than the separation between the lowest vibration levels. In this context, the only cage distribution that has to be considered is that of the singlet vibration ground state. Thanks to this spherical simplification, it is shown that, for thermal energies below the electronic ground state width:\newline
- a Schottky like anomaly develops in the specific heat, that evidences the decrease in the 4$f$ energy and entropy. \newline
- the magnetic susceptibility is reduced with respect to the degenerate case.\newline
- some volume anomaly should occur, reflecting the expansion of the rare-earth distribution inside the cage.\newline
- the centrifugal term reduces the rattling frequency of the guest, with resulting softening of low energy phonons.\newline
This model also shows that the above conclusions are not dependent on the precise shape of the cage potential. Indeed, the two dominant energy scales in the problem are:\newline
- the separation $\Delta E$ between the lowest lying and first excited vibration levels, which essentially depend on the size of the cage,\newline
- the smaller energy width of the electronic ground-state ($\approx 2 \beta$), that defines the amplitude and temperature range of the phenomena resulting from the split CEF ground state.\newline
This is encouraging as regards the description, in the paramagnetic range, of the macroscopic properties of a real cage system. Once defined the central CEF ground state, the rare-earth distribution in the lowest vibrational state is required for computing the specific heat, entropy, magnetic susceptibility, ... . In this purpose, the detail of the cage shape and associated potential well can be ignored. Indeed, the wave function for the singlet vibration ground state will be very similar for all highly symmetrical shapes of the potential well : harmonic well, spherical or cubic box, etc.. What actually matters is the width of the rare-earth distribution and the magnitude of the quadrupole-displacement coupling constants.
In this way, one may explain, by a pure crystal field mechanism, paramagnetic anomalies otherwise ascribed to exchange couplings, such as magnetic correlations or Kondo lattice related effects.\newline
An interesting perspective, at the price of an increased difficulty, is the description of paramagnetic properties under higher magnetic fields. The contrast in the local magnetic susceptibilities inside the cage should result in an anisotropic potential well: cage quadrupoles should emerge with an associated magnetostriction and influence on the magnetic anisotropy. Another, completely unexplored, aspect is the influence of a distributed crystal field on the ordering properties of a rare-earth compound. It is hoped that the present work will encourage investigations in these directions.\newline

\textbf{Acknowlegments}\newline
The author would like to thank Dr Rose-Marie Gal\'era and Dr Julien Robert for their critical reading of the manuscript.

\appendix
\section{Dispersion relations for a linear chain of elastic cages}
\label{RelDisp}
The chain, of period $d$, is defined in Fig. \ref{CageDisp}, through the sketch of the involved masses and springs. One starts by writing the classical equations of motion for the guest (mass $m$) and the two halves (masses $M/2$) of the cage, subjected to the springs restoring forces. Then, imposing the form of the displacements as imaginary exponentials describing propagating waves with angular frequency $\omega$ and wave-vector $q$, one gets the secular equation:
\begin{widetext}

\begin{equation}
\left[ {{\Omega _0}^2 (1 - \cos(q d)) + 2 {\omega ^2}\frac{{(1 + \alpha ) {\omega _0}^2 - {\omega ^2}}}{{{\omega^2} - {\omega_0}^2}}} \right]\left[ {{\Omega _0}^2\ {(1 + \cos (q d))} +2( {\Omega _1}^2 + \alpha \; {\omega _0}^2 - {\omega ^2}} )\right] = {\Omega_0}^4 {\sin(q d)}^2
\end{equation}

where $ \alpha = \frac{m}{M} $,  $ {\omega_0}^2 = \frac{2 k_0}{m} $, $ {\Omega_0}^2  = \frac{4 K_0}{M} $,  ${\Omega_1}^2 = \frac{4 K_1}{M} $.

Further parametrizing, introducing the reduced variable $x=\omega / \omega_0$, frequencies $\rho_0=\Omega_0 / \omega_0$ and $\rho_1=\Omega_1 / \omega_0$, replacing ${\sin(q d)}^2$ with $1-{\cos(q d)}^2=(1-{\cos(q d)})(1+{\cos(q d)})$, some simplifications yield:\newline
\begin{equation}
{\rho _0}^2 (1 - \cos (q d)) ({\rho_1}^2 + \alpha  - {x^2})({x^2} - 1) + {x^2}(\alpha  - {x^2} + 1)\left[ {\rho_0}^2(1 + \cos (q d)) + 2(\alpha +  {\rho_1}^2 - {x^2}) \right] = 0
\end{equation}
Isolating $\cos(q d)$, the general dispersion relation can be expressed as:\newline
\begin{equation}
\cos (q d) =1- 2 \frac{x^2}{\rho_0^2} \frac{\left(\alpha -x^2+1\right) \left(\alpha +\rho _1^2-x^2\right)} {\alpha +\rho_1^2 \left(1-x^2\right)}+ \frac{2 x^2 \left(x^2-\alpha -1\right)} {\alpha +\rho_1^2 \left(1-x^2\right)} 
\end{equation}
In the limit of an infinitely soft cage (${\rho_1} = 0$), the dispersion relation reads as:\newline
\begin{equation}
\cos (q d) =1- 2 \frac{x^2}{\alpha \rho_0^2} \left(\alpha -x^2+1\right) \left(\alpha-x^2\right)+ \frac{2 x^2 } {\alpha} \left(x^2-\alpha -1\right)
\end{equation}
whereas, for an infinitely rigid cage (${\rho_1} \rightarrow +\infty$), the dispersion relation simplifies to:
\begin{equation} 
\cos (q d) =1- \frac{2}{\rho_0^2} \left(1+\frac{\alpha}{ 1-x^2}\right) x^2
\end{equation}
 
\end{widetext}

\section{Solving the radial Schr\"{o}dinger equation}
\label{NumSchroed}
The spherical symmetry allows to use the methods, easily found in quantum mechanics textbooks, developed for the quantum description of atoms.
The angular part of the wave functions are the spherical harmonics $Y_{l}^{m}(\theta,\varphi)$, whereas the radial part $R_{n,l}(\rho)$ and energies $E_{n,l}$ require solving a differential equation specific to the considered potential, here $V(\rho, T)$. It is convenient to introduce the radial function $u_{n,l}$:
\begin{equation}
u_{n,l}(\rho)= \rho\;R_{n,l}(\rho)\nonumber
\end{equation} 
For $\rho<1$, in the case of the above defined potential well, the sought after $u_{n,l}(\rho)$ functions have to satisfy the differential equation: 
\begin{equation}
\label{SchroeRad}
\frac{d^2}{d{\rho^2}} u_{n,l} = (\frac{l(l + 1)}{\rho^2} + V(\rho, T) - E_{n,l}) \, u_{n,l}
\end{equation}
where $l$ is the orbital quantum number and $E_{n,l}$ the energy associated with $u_{n,l}$ (all energies are in $e.u.$ unit, as defined in Section \ref{SpherWell}). 
At the cage limit $\rho = 1$, the infinite barrier imposes $u_{n,l}(1)=0$.  At the origin, $u_{n,l}(\rho)$ has also to cancel as, otherwise, one would get a diverging radial wave function $R_{n,l}$ for $\rho$ going to zero. In the general case, the only practical option is a numerical treatment of Eq. (\ref{SchroeRad}).
For a given value of $l$ and numerically evaluating $V(\rho, T)$, one can iteratively refine a numerical solution, which yields the radial wave function $R_{n,l}$ and corresponding energy $E_{n,l}$. The calculations presented in this paper are realized using a simple Numerov type method, the 0 to 1 radial interval being divided into 2000 equal segments. The convergence is tested by checking the stability of the $E_{n,l}$ eigenvalue, which is required to vary less than 10$^{-8}$ $e.u.$ between successive iterations. Obviously, the obtained $R_{n,l}$ need to be normalized so that $4 \pi \rho^2 |R_{n,l}(\rho)|^2$ can be interpreted as the radial probability of presence of the nucleus.




\begin{thebibliography}{48}%
\makeatletter
\providecommand \@ifxundefined [1]{%
 \@ifx{#1\undefined}
}%
\providecommand \@ifnum [1]{%
 \ifnum #1\expandafter \@firstoftwo
 \else \expandafter \@secondoftwo
 \fi
}%
\providecommand \@ifx [1]{%
 \ifx #1\expandafter \@firstoftwo
 \else \expandafter \@secondoftwo
 \fi
}%
\providecommand \natexlab [1]{#1}%
\providecommand \enquote  [1]{``#1''}%
\providecommand \bibnamefont  [1]{#1}%
\providecommand \bibfnamefont [1]{#1}%
\providecommand \citenamefont [1]{#1}%
\providecommand \href@noop [0]{\@secondoftwo}%
\providecommand \href [0]{\begingroup \@sanitize@url \@href}%
\providecommand \@href[1]{\@@startlink{#1}\@@href}%
\providecommand \@@href[1]{\endgroup#1\@@endlink}%
\providecommand \@sanitize@url [0]{\catcode `\\12\catcode `\$12\catcode
  `\&12\catcode `\#12\catcode `\^12\catcode `\_12\catcode `\%12\relax}%
\providecommand \@@startlink[1]{}%
\providecommand \@@endlink[0]{}%
\providecommand \url  [0]{\begingroup\@sanitize@url \@url }%
\providecommand \@url [1]{\endgroup\@href {#1}{\urlprefix }}%
\providecommand \urlprefix  [0]{URL }%
\providecommand \Eprint [0]{\href }%
\providecommand \doibase [0]{http://dx.doi.org/}%
\providecommand \selectlanguage [0]{\@gobble}%
\providecommand \bibinfo  [0]{\@secondoftwo}%
\providecommand \bibfield  [0]{\@secondoftwo}%
\providecommand \translation [1]{[#1]}%
\providecommand \BibitemOpen [0]{}%
\providecommand \bibitemStop [0]{}%
\providecommand \bibitemNoStop [0]{.\EOS\space}%
\providecommand \EOS [0]{\spacefactor3000\relax}%
\providecommand \BibitemShut  [1]{\csname bibitem#1\endcsname}%
\let\auto@bib@innerbib\@empty
\bibitem [{\citenamefont {Jeitschko}\ and\ \citenamefont
  {Braun}(1977)}]{Jeitschko1977}%
  \BibitemOpen
  \bibfield  {author} {\bibinfo {author} {\bibfnamefont {W.}~\bibnamefont
  {Jeitschko}}\ and\ \bibinfo {author} {\bibfnamefont {D.}~\bibnamefont
  {Braun}},\ }\href {\doibase 10.1107/S056774087701108X} {\bibfield  {journal}
  {\bibinfo  {journal} {Acta Crystallographica Section B}\ }\textbf {\bibinfo
  {volume} {33}},\ \bibinfo {pages} {3401} (\bibinfo {year}
  {1977})}\BibitemShut {NoStop}%
\bibitem [{\citenamefont {Bauer}\ \emph {et~al.}(2002)\citenamefont {Bauer},
  \citenamefont {Frederick}, \citenamefont {Ho}, \citenamefont {Zapf},\ and\
  \citenamefont {Maple}}]{Bauer2002}%
  \BibitemOpen
  \bibfield  {author} {\bibinfo {author} {\bibfnamefont {E.~D.}\ \bibnamefont
  {Bauer}}, \bibinfo {author} {\bibfnamefont {N.~A.}\ \bibnamefont
  {Frederick}}, \bibinfo {author} {\bibfnamefont {P.-C.}\ \bibnamefont {Ho}},
  \bibinfo {author} {\bibfnamefont {V.~S.}\ \bibnamefont {Zapf}}, \ and\
  \bibinfo {author} {\bibfnamefont {M.~B.}\ \bibnamefont {Maple}},\ }\href
  {\doibase 10.1103/PhysRevB.65.100506} {\bibfield  {journal} {\bibinfo
  {journal} {Phys. Rev. B}\ }\textbf {\bibinfo {volume} {65}},\ \bibinfo
  {pages} {100506} (\bibinfo {year} {2002})}\BibitemShut {NoStop}%
\bibitem [{\citenamefont {Iwasa}\ \emph {et~al.}(2005)\citenamefont {Iwasa},
  \citenamefont {Hao}, \citenamefont {Kuwahara}, \citenamefont {Kohgi},
  \citenamefont {Saha}, \citenamefont {Sugawara}, \citenamefont {Aoki},
  \citenamefont {Sato}, \citenamefont {Tayama},\ and\ \citenamefont
  {Sakakibara}}]{Iwasa2005}%
  \BibitemOpen
  \bibfield  {author} {\bibinfo {author} {\bibfnamefont {K.}~\bibnamefont
  {Iwasa}}, \bibinfo {author} {\bibfnamefont {L.}~\bibnamefont {Hao}}, \bibinfo
  {author} {\bibfnamefont {K.}~\bibnamefont {Kuwahara}}, \bibinfo {author}
  {\bibfnamefont {M.}~\bibnamefont {Kohgi}}, \bibinfo {author} {\bibfnamefont
  {S.~R.}\ \bibnamefont {Saha}}, \bibinfo {author} {\bibfnamefont
  {H.}~\bibnamefont {Sugawara}}, \bibinfo {author} {\bibfnamefont
  {Y.}~\bibnamefont {Aoki}}, \bibinfo {author} {\bibfnamefont {H.}~\bibnamefont
  {Sato}}, \bibinfo {author} {\bibfnamefont {T.}~\bibnamefont {Tayama}}, \ and\
  \bibinfo {author} {\bibfnamefont {T.}~\bibnamefont {Sakakibara}},\ }\href
  {\doibase 10.1103/PhysRevB.72.024414} {\bibfield  {journal} {\bibinfo
  {journal} {Phys. Rev. B}\ }\textbf {\bibinfo {volume} {72}},\ \bibinfo
  {pages} {024414} (\bibinfo {year} {2005})}\BibitemShut {NoStop}%
\bibitem [{\citenamefont {Keller}\ \emph {et~al.}(2001)\citenamefont {Keller},
  \citenamefont {Fischer}, \citenamefont {Herrmannsd{\"o}rfer}, \citenamefont
  {D{\"o}nni}, \citenamefont {Sugawara}, \citenamefont {Matsuda}, \citenamefont
  {Abe}, \citenamefont {Aoki},\ and\ \citenamefont {Sato}}]{Keller2001}%
  \BibitemOpen
  \bibfield  {author} {\bibinfo {author} {\bibfnamefont {L.}~\bibnamefont
  {Keller}}, \bibinfo {author} {\bibfnamefont {P.}~\bibnamefont {Fischer}},
  \bibinfo {author} {\bibfnamefont {T.}~\bibnamefont {Herrmannsd{\"o}rfer}},
  \bibinfo {author} {\bibfnamefont {A.}~\bibnamefont {D{\"o}nni}}, \bibinfo
  {author} {\bibfnamefont {H.}~\bibnamefont {Sugawara}}, \bibinfo {author}
  {\bibfnamefont {T.}~\bibnamefont {Matsuda}}, \bibinfo {author} {\bibfnamefont
  {K.}~\bibnamefont {Abe}}, \bibinfo {author} {\bibfnamefont {Y.}~\bibnamefont
  {Aoki}}, \ and\ \bibinfo {author} {\bibfnamefont {H.}~\bibnamefont {Sato}},\
  }\href {\doibase http://dx.doi.org/10.1016/S0925-8388(01)01163-X} {\bibfield
  {journal} {\bibinfo  {journal} {Journal of Alloys and Compounds}\ }\textbf
  {\bibinfo {volume} {323--324}},\ \bibinfo {pages} {516 } (\bibinfo {year}
  {2001})},\ \bibinfo {note} {proceedings of the 4th International Conference
  on f-Elements}\BibitemShut {NoStop}%
\bibitem [{\citenamefont {Effantin}\ \emph {et~al.}(1985)\citenamefont
  {Effantin}, \citenamefont {Rossat-Mignod}, \citenamefont {Burlet},
  \citenamefont {Bartholin}, \citenamefont {Kunii},\ and\ \citenamefont
  {Kasuya}}]{Effantin1985}%
  \BibitemOpen
  \bibfield  {author} {\bibinfo {author} {\bibfnamefont {J.}~\bibnamefont
  {Effantin}}, \bibinfo {author} {\bibfnamefont {J.}~\bibnamefont
  {Rossat-Mignod}}, \bibinfo {author} {\bibfnamefont {P.}~\bibnamefont
  {Burlet}}, \bibinfo {author} {\bibfnamefont {H.}~\bibnamefont {Bartholin}},
  \bibinfo {author} {\bibfnamefont {S.}~\bibnamefont {Kunii}}, \ and\ \bibinfo
  {author} {\bibfnamefont {T.}~\bibnamefont {Kasuya}},\ }\href@noop {}
  {\bibfield  {journal} {\bibinfo  {journal} {J. Magn. Magn. Mater.}\ }\textbf
  {\bibinfo {volume} {47-48}},\ \bibinfo {pages} {145} (\bibinfo {year}
  {1985})}\BibitemShut {NoStop}%
\bibitem [{\citenamefont {Amara}\ and\ \citenamefont
  {Gal\'era}(2012)}]{Amara2012}%
  \BibitemOpen
  \bibfield  {author} {\bibinfo {author} {\bibfnamefont {M.}~\bibnamefont
  {Amara}}\ and\ \bibinfo {author} {\bibfnamefont {R.-M.}\ \bibnamefont
  {Gal\'era}},\ }\href {\doibase 10.1103/PhysRevLett.108.026402} {\bibfield
  {journal} {\bibinfo  {journal} {Phys. Rev. Lett.}\ }\textbf {\bibinfo
  {volume} {108}},\ \bibinfo {pages} {026402} (\bibinfo {year}
  {2012})}\BibitemShut {NoStop}%
\bibitem [{\citenamefont {Bethe}(1929)}]{bethe1929}%
  \BibitemOpen
  \bibfield  {author} {\bibinfo {author} {\bibfnamefont {H.}~\bibnamefont
  {Bethe}},\ }\href@noop {} {\bibfield  {journal} {\bibinfo  {journal} {Annalen
  der Physik}\ }\textbf {\bibinfo {volume} {3}},\ \bibinfo {pages} {133}
  (\bibinfo {year} {1929})}\BibitemShut {NoStop}%
\bibitem [{\citenamefont {Park}\ \emph {et~al.}(2008)\citenamefont {Park},
  \citenamefont {Adroja}, \citenamefont {McEwen}, \citenamefont {Kohgi},\ and\
  \citenamefont {Iwasa}}]{Park2008}%
  \BibitemOpen
  \bibfield  {author} {\bibinfo {author} {\bibfnamefont {J.-G.}\ \bibnamefont
  {Park}}, \bibinfo {author} {\bibfnamefont {D.~T.}\ \bibnamefont {Adroja}},
  \bibinfo {author} {\bibfnamefont {K.~A.}\ \bibnamefont {McEwen}}, \bibinfo
  {author} {\bibfnamefont {M.}~\bibnamefont {Kohgi}}, \ and\ \bibinfo {author}
  {\bibfnamefont {K.}~\bibnamefont {Iwasa}},\ }\href {\doibase
  10.1103/PhysRevB.77.085102} {\bibfield  {journal} {\bibinfo  {journal} {Phys.
  Rev. B}\ }\textbf {\bibinfo {volume} {77}},\ \bibinfo {pages} {085102}
  (\bibinfo {year} {2008})}\BibitemShut {NoStop}%
\bibitem [{\citenamefont {Iwasa}\ \emph {et~al.}(2009)\citenamefont {Iwasa},
  \citenamefont {Saito}, \citenamefont {Murakami},\ and\ \citenamefont
  {Sugawara}}]{Iwasa2009}%
  \BibitemOpen
  \bibfield  {author} {\bibinfo {author} {\bibfnamefont {K.}~\bibnamefont
  {Iwasa}}, \bibinfo {author} {\bibfnamefont {K.}~\bibnamefont {Saito}},
  \bibinfo {author} {\bibfnamefont {Y.}~\bibnamefont {Murakami}}, \ and\
  \bibinfo {author} {\bibfnamefont {H.}~\bibnamefont {Sugawara}},\ }\href
  {\doibase 10.1103/PhysRevB.79.235113} {\bibfield  {journal} {\bibinfo
  {journal} {Phys. Rev. B}\ }\textbf {\bibinfo {volume} {79}},\ \bibinfo
  {pages} {235113} (\bibinfo {year} {2009})}\BibitemShut {NoStop}%
\bibitem [{\citenamefont {Zirngiebl}\ \emph {et~al.}(1984)\citenamefont
  {Zirngiebl}, \citenamefont {Hillebrands}, \citenamefont {Blumenr\"oder},
  \citenamefont {G\"untherodt}, \citenamefont {Loewenhaupt}, \citenamefont
  {Carpenter}, \citenamefont {Winzer},\ and\ \citenamefont
  {Fisk}}]{Zirngiebl1984}%
  \BibitemOpen
  \bibfield  {author} {\bibinfo {author} {\bibfnamefont {E.}~\bibnamefont
  {Zirngiebl}}, \bibinfo {author} {\bibfnamefont {B.}~\bibnamefont
  {Hillebrands}}, \bibinfo {author} {\bibfnamefont {S.}~\bibnamefont
  {Blumenr\"oder}}, \bibinfo {author} {\bibfnamefont {G.}~\bibnamefont
  {G\"untherodt}}, \bibinfo {author} {\bibfnamefont {M.}~\bibnamefont
  {Loewenhaupt}}, \bibinfo {author} {\bibfnamefont {J.~M.}\ \bibnamefont
  {Carpenter}}, \bibinfo {author} {\bibfnamefont {K.}~\bibnamefont {Winzer}}, \
  and\ \bibinfo {author} {\bibfnamefont {Z.}~\bibnamefont {Fisk}},\ }\href
  {\doibase 10.1103/PhysRevB.30.4052} {\bibfield  {journal} {\bibinfo
  {journal} {Phys. Rev. B}\ }\textbf {\bibinfo {volume} {30}},\ \bibinfo
  {pages} {4052} (\bibinfo {year} {1984})}\BibitemShut {NoStop}%
\bibitem [{\citenamefont {Loewenhaupt}\ and\ \citenamefont
  {Prager}(1986)}]{Loewenhaupt1986}%
  \BibitemOpen
  \bibfield  {author} {\bibinfo {author} {\bibfnamefont {M.}~\bibnamefont
  {Loewenhaupt}}\ and\ \bibinfo {author} {\bibfnamefont {M.}~\bibnamefont
  {Prager}},\ }\href {\doibase 10.1007/BF01323430} {\bibfield  {journal}
  {\bibinfo  {journal} {Zeitschrift f{\"u}r Physik B Condensed Matter}\
  }\textbf {\bibinfo {volume} {62}},\ \bibinfo {pages} {195} (\bibinfo {year}
  {1986})}\BibitemShut {NoStop}%
\bibitem [{\citenamefont {Fujita}\ \emph {et~al.}(1980)\citenamefont {Fujita},
  \citenamefont {Suzuki}, \citenamefont {Komatsubara}, \citenamefont {Kunii},
  \citenamefont {Kasuya},\ and\ \citenamefont {Ohtsuka}}]{Fujita1980}%
  \BibitemOpen
  \bibfield  {author} {\bibinfo {author} {\bibfnamefont {T.}~\bibnamefont
  {Fujita}}, \bibinfo {author} {\bibfnamefont {M.}~\bibnamefont {Suzuki}},
  \bibinfo {author} {\bibfnamefont {T.}~\bibnamefont {Komatsubara}}, \bibinfo
  {author} {\bibfnamefont {S.}~\bibnamefont {Kunii}}, \bibinfo {author}
  {\bibfnamefont {T.}~\bibnamefont {Kasuya}}, \ and\ \bibinfo {author}
  {\bibfnamefont {T.}~\bibnamefont {Ohtsuka}},\ }\href {\doibase DOI:
  10.1016/0038-1098(80)90900-X} {\bibfield  {journal} {\bibinfo  {journal}
  {Solid State Communications}\ }\textbf {\bibinfo {volume} {35}},\ \bibinfo
  {pages} {569 } (\bibinfo {year} {1980})}\BibitemShut {NoStop}%
\bibitem [{\citenamefont {{P}eysson}\ \emph {et~al.}(1986)\citenamefont
  {{P}eysson}, \citenamefont {{A}yache}, \citenamefont {{R}ossat {M}ignod},
  \citenamefont {{K}unii},\ and\ \citenamefont {{K}asuya}}]{PEYSSON1986}%
  \BibitemOpen
  \bibfield  {author} {\bibinfo {author} {\bibfnamefont {Y.}~\bibnamefont
  {{P}eysson}}, \bibinfo {author} {\bibfnamefont {C.}~\bibnamefont {{A}yache}},
  \bibinfo {author} {\bibfnamefont {J.}~\bibnamefont {{R}ossat {M}ignod}},
  \bibinfo {author} {\bibfnamefont {S.}~\bibnamefont {{K}unii}}, \ and\
  \bibinfo {author} {\bibfnamefont {T.}~\bibnamefont {{K}asuya}},\ }\href
  {\doibase 10.1051/jphys:01986004701011300} {\bibfield  {journal} {\bibinfo
  {journal} {{J}ournal de {P}hysique}\ }\textbf {\bibinfo {volume} {47}},\
  \bibinfo {pages} {113} (\bibinfo {year} {1986})}\BibitemShut {NoStop}%
\bibitem [{\citenamefont {Lee}\ \emph {et~al.}(1970)\citenamefont {Lee},
  \citenamefont {Bachmann}, \citenamefont {Geballe},\ and\ \citenamefont
  {Maita}}]{Lee1970}%
  \BibitemOpen
  \bibfield  {author} {\bibinfo {author} {\bibfnamefont {K.~N.}\ \bibnamefont
  {Lee}}, \bibinfo {author} {\bibfnamefont {R.}~\bibnamefont {Bachmann}},
  \bibinfo {author} {\bibfnamefont {T.~H.}\ \bibnamefont {Geballe}}, \ and\
  \bibinfo {author} {\bibfnamefont {J.~P.}\ \bibnamefont {Maita}},\ }\href
  {\doibase 10.1103/PhysRevB.2.4580} {\bibfield  {journal} {\bibinfo  {journal}
  {Phys. Rev. B}\ }\textbf {\bibinfo {volume} {2}},\ \bibinfo {pages} {4580}
  (\bibinfo {year} {1970})}\BibitemShut {NoStop}%
\bibitem [{\citenamefont {Westrum~Jr}\ \emph {et~al.}(1965)\citenamefont
  {Westrum~Jr}, \citenamefont {Clever}, \citenamefont {Andrews},\ and\
  \citenamefont {Feick}}]{westrum1964}%
  \BibitemOpen
  \bibfield  {author} {\bibinfo {author} {\bibfnamefont {E.}~\bibnamefont
  {Westrum~Jr}}, \bibinfo {author} {\bibfnamefont {H.}~\bibnamefont {Clever}},
  \bibinfo {author} {\bibfnamefont {J.}~\bibnamefont {Andrews}}, \ and\
  \bibinfo {author} {\bibfnamefont {G.}~\bibnamefont {Feick}}\ }(\bibinfo
  {publisher} {Gordon and Breach},\ \bibinfo {year} {1965})\ p.\ \bibinfo
  {pages} {597}\BibitemShut {NoStop}%
\bibitem [{\citenamefont {Loewenhaupt}\ \emph {et~al.}(1985)\citenamefont
  {Loewenhaupt}, \citenamefont {Carpenter},\ and\ \citenamefont
  {Loong}}]{Loewenhaupt1985}%
  \BibitemOpen
  \bibfield  {author} {\bibinfo {author} {\bibfnamefont {M.}~\bibnamefont
  {Loewenhaupt}}, \bibinfo {author} {\bibfnamefont {J.}~\bibnamefont
  {Carpenter}}, \ and\ \bibinfo {author} {\bibfnamefont {C.-K.}\ \bibnamefont
  {Loong}},\ }\href {\doibase 10.1016/0304-8853(85)90270-7} {\bibfield
  {journal} {\bibinfo  {journal} {Journal of Magnetism and Magnetic Materials}\
  }\textbf {\bibinfo {volume} {52}},\ \bibinfo {pages} {245 } (\bibinfo {year}
  {1985})}\BibitemShut {NoStop}%
\bibitem [{\citenamefont {Ho}\ \emph {et~al.}(2005)\citenamefont {Ho},
  \citenamefont {Yuhasz}, \citenamefont {Butch}, \citenamefont {Frederick},
  \citenamefont {Sayles}, \citenamefont {Jeffries}, \citenamefont {Maple},
  \citenamefont {Betts}, \citenamefont {Lacerda}, \citenamefont {Rogl},\ and\
  \citenamefont {Giester}}]{Ho2005}%
  \BibitemOpen
  \bibfield  {author} {\bibinfo {author} {\bibfnamefont {P.-C.}\ \bibnamefont
  {Ho}}, \bibinfo {author} {\bibfnamefont {W.~M.}\ \bibnamefont {Yuhasz}},
  \bibinfo {author} {\bibfnamefont {N.~P.}\ \bibnamefont {Butch}}, \bibinfo
  {author} {\bibfnamefont {N.~A.}\ \bibnamefont {Frederick}}, \bibinfo {author}
  {\bibfnamefont {T.~A.}\ \bibnamefont {Sayles}}, \bibinfo {author}
  {\bibfnamefont {J.~R.}\ \bibnamefont {Jeffries}}, \bibinfo {author}
  {\bibfnamefont {M.~B.}\ \bibnamefont {Maple}}, \bibinfo {author}
  {\bibfnamefont {J.~B.}\ \bibnamefont {Betts}}, \bibinfo {author}
  {\bibfnamefont {A.~H.}\ \bibnamefont {Lacerda}}, \bibinfo {author}
  {\bibfnamefont {P.}~\bibnamefont {Rogl}}, \ and\ \bibinfo {author}
  {\bibfnamefont {G.}~\bibnamefont {Giester}},\ }\href {\doibase
  10.1103/PhysRevB.72.094410} {\bibfield  {journal} {\bibinfo  {journal} {Phys.
  Rev. B}\ }\textbf {\bibinfo {volume} {72}},\ \bibinfo {pages} {094410}
  (\bibinfo {year} {2005})}\BibitemShut {NoStop}%
\bibitem [{\citenamefont {Yuhasz}\ \emph {et~al.}(2006)\citenamefont {Yuhasz},
  \citenamefont {Butch}, \citenamefont {Sayles}, \citenamefont {Ho},
  \citenamefont {Jeffries}, \citenamefont {Yanagisawa}, \citenamefont
  {Frederick}, \citenamefont {Maple}, \citenamefont {Henkie}, \citenamefont
  {Pietraszko}, \citenamefont {McCall}, \citenamefont {McElfresh},\ and\
  \citenamefont {Fluss}}]{Yuhasz2006}%
  \BibitemOpen
  \bibfield  {author} {\bibinfo {author} {\bibfnamefont {W.~M.}\ \bibnamefont
  {Yuhasz}}, \bibinfo {author} {\bibfnamefont {N.~P.}\ \bibnamefont {Butch}},
  \bibinfo {author} {\bibfnamefont {T.~A.}\ \bibnamefont {Sayles}}, \bibinfo
  {author} {\bibfnamefont {P.-C.}\ \bibnamefont {Ho}}, \bibinfo {author}
  {\bibfnamefont {J.~R.}\ \bibnamefont {Jeffries}}, \bibinfo {author}
  {\bibfnamefont {T.}~\bibnamefont {Yanagisawa}}, \bibinfo {author}
  {\bibfnamefont {N.~A.}\ \bibnamefont {Frederick}}, \bibinfo {author}
  {\bibfnamefont {M.~B.}\ \bibnamefont {Maple}}, \bibinfo {author}
  {\bibfnamefont {Z.}~\bibnamefont {Henkie}}, \bibinfo {author} {\bibfnamefont
  {A.}~\bibnamefont {Pietraszko}}, \bibinfo {author} {\bibfnamefont {S.~K.}\
  \bibnamefont {McCall}}, \bibinfo {author} {\bibfnamefont {M.~W.}\
  \bibnamefont {McElfresh}}, \ and\ \bibinfo {author} {\bibfnamefont {M.~J.}\
  \bibnamefont {Fluss}},\ }\href {\doibase 10.1103/PhysRevB.73.144409}
  {\bibfield  {journal} {\bibinfo  {journal} {Phys. Rev. B}\ }\textbf {\bibinfo
  {volume} {73}},\ \bibinfo {pages} {144409} (\bibinfo {year}
  {2006})}\BibitemShut {NoStop}%
\bibitem [{\citenamefont {Cichorek}\ \emph {et~al.}(2014)\citenamefont
  {Cichorek}, \citenamefont {Rudenko}, \citenamefont {Wisniewski},
  \citenamefont {Wawryk}, \citenamefont {Kepinski}, \citenamefont
  {Pietraszko},\ and\ \citenamefont {Henkie}}]{Cichorek2014}%
  \BibitemOpen
  \bibfield  {author} {\bibinfo {author} {\bibfnamefont {T.}~\bibnamefont
  {Cichorek}}, \bibinfo {author} {\bibfnamefont {A.}~\bibnamefont {Rudenko}},
  \bibinfo {author} {\bibfnamefont {P.}~\bibnamefont {Wisniewski}}, \bibinfo
  {author} {\bibfnamefont {R.}~\bibnamefont {Wawryk}}, \bibinfo {author}
  {\bibfnamefont {L.}~\bibnamefont {Kepinski}}, \bibinfo {author}
  {\bibfnamefont {A.}~\bibnamefont {Pietraszko}}, \ and\ \bibinfo {author}
  {\bibfnamefont {Z.}~\bibnamefont {Henkie}},\ }\href {\doibase
  10.1103/PhysRevB.90.195123} {\bibfield  {journal} {\bibinfo  {journal} {Phys.
  Rev. B}\ }\textbf {\bibinfo {volume} {90}},\ \bibinfo {pages} {195123}
  (\bibinfo {year} {2014})}\BibitemShut {NoStop}%
\bibitem [{\citenamefont {{van Vleck}}(1939)}]{VanVleck1939}%
  \BibitemOpen
  \bibfield  {author} {\bibinfo {author} {\bibfnamefont {J.~H.}\ \bibnamefont
  {{van Vleck}}},\ }\href {\doibase 10.1063/1.1750327} {\bibfield  {journal}
  {\bibinfo  {journal} {The Journal of Chemical Physics}\ }\textbf {\bibinfo
  {volume} {7}},\ \bibinfo {pages} {72} (\bibinfo {year} {1939})}\BibitemShut
  {NoStop}%
\bibitem [{\citenamefont {Opik}\ and\ \citenamefont {Pryce}(1957)}]{Opik1957}%
  \BibitemOpen
  \bibfield  {author} {\bibinfo {author} {\bibfnamefont {U.}~\bibnamefont
  {Opik}}\ and\ \bibinfo {author} {\bibfnamefont {M.~H.~L.}\ \bibnamefont
  {Pryce}},\ }\href {\doibase 10.1098/rspa.1957.0010} {\bibfield  {journal}
  {\bibinfo  {journal} {Proceedings of the Royal Society of London. Series A.
  Mathematical and Physical Sciences}\ }\textbf {\bibinfo {volume} {238}},\
  \bibinfo {pages} {425} (\bibinfo {year} {1957})}\BibitemShut {NoStop}%
\bibitem [{\citenamefont {Gehring}\ and\ \citenamefont
  {Gehring}(1975)}]{Gehring1975}%
  \BibitemOpen
  \bibfield  {author} {\bibinfo {author} {\bibfnamefont {G.}~\bibnamefont
  {Gehring}}\ and\ \bibinfo {author} {\bibfnamefont {K.}~\bibnamefont
  {Gehring}},\ }\href@noop {} {\bibfield  {journal} {\bibinfo  {journal} {Rep.
  Prog. Phys.}\ }\textbf {\bibinfo {volume} {38}},\ \bibinfo {pages} {1}
  (\bibinfo {year} {1975})}\BibitemShut {NoStop}%
\bibitem [{\citenamefont {Chernyshov}\ \emph {et~al.}(1997)\citenamefont
  {Chernyshov}, \citenamefont {Smirnov}, \citenamefont {Menschikova},
  \citenamefont {Mirgorodsky},\ and\ \citenamefont {Trounov}}]{CHERNYSHOV1997}%
  \BibitemOpen
  \bibfield  {author} {\bibinfo {author} {\bibfnamefont {D.}~\bibnamefont
  {Chernyshov}}, \bibinfo {author} {\bibfnamefont {M.}~\bibnamefont {Smirnov}},
  \bibinfo {author} {\bibfnamefont {A.}~\bibnamefont {Menschikova}}, \bibinfo
  {author} {\bibfnamefont {A.}~\bibnamefont {Mirgorodsky}}, \ and\ \bibinfo
  {author} {\bibfnamefont {V.}~\bibnamefont {Trounov}},\ }\href {\doibase
  http://dx.doi.org/10.1016/S0921-4526(96)00931-3} {\bibfield  {journal}
  {\bibinfo  {journal} {Physica B: Condensed Matter}\ }\textbf {\bibinfo
  {volume} {234}},\ \bibinfo {pages} {146 } (\bibinfo {year} {1997})},\
  \bibinfo {note} {proceedings of the First European Conference on Neutron
  Scattering}\BibitemShut {NoStop}%
\bibitem [{\citenamefont {Takahashi}\ \emph {et~al.}(1999)\citenamefont
  {Takahashi}, \citenamefont {Ohshima}, \citenamefont {Okamura}, \citenamefont
  {Otani},\ and\ \citenamefont {Tanaka}}]{Takahashi1999}%
  \BibitemOpen
  \bibfield  {author} {\bibinfo {author} {\bibfnamefont {Y.}~\bibnamefont
  {Takahashi}}, \bibinfo {author} {\bibfnamefont {K.}~\bibnamefont {Ohshima}},
  \bibinfo {author} {\bibfnamefont {F.~P.}\ \bibnamefont {Okamura}}, \bibinfo
  {author} {\bibfnamefont {S.}~\bibnamefont {Otani}}, \ and\ \bibinfo {author}
  {\bibfnamefont {T.}~\bibnamefont {Tanaka}},\ } {\bibfield  {journal} {\bibinfo  {journal} {Journal of
  the Physical Society of Japan}\ }\textbf {\bibinfo {volume} {68}},\ \bibinfo
  {pages} {2304} (\bibinfo {year} {1999})}\BibitemShut {NoStop}%
\bibitem [{\citenamefont {Schnelle}\ \emph {et~al.}(2008)\citenamefont
  {Schnelle}, \citenamefont {Leithe-Jasper}, \citenamefont {Rosner},
  \citenamefont {Cardoso-Gil}, \citenamefont {Gumeniuk}, \citenamefont {Trots},
  \citenamefont {Mydosh},\ and\ \citenamefont {Grin}}]{Schnelle2008}%
  \BibitemOpen
  \bibfield  {author} {\bibinfo {author} {\bibfnamefont {W.}~\bibnamefont
  {Schnelle}}, \bibinfo {author} {\bibfnamefont {A.}~\bibnamefont
  {Leithe-Jasper}}, \bibinfo {author} {\bibfnamefont {H.}~\bibnamefont
  {Rosner}}, \bibinfo {author} {\bibfnamefont {R.}~\bibnamefont {Cardoso-Gil}},
  \bibinfo {author} {\bibfnamefont {R.}~\bibnamefont {Gumeniuk}}, \bibinfo
  {author} {\bibfnamefont {D.}~\bibnamefont {Trots}}, \bibinfo {author}
  {\bibfnamefont {J.~A.}\ \bibnamefont {Mydosh}}, \ and\ \bibinfo {author}
  {\bibfnamefont {Y.}~\bibnamefont {Grin}},\ }\href {\doibase
  10.1103/PhysRevB.77.094421} {\bibfield  {journal} {\bibinfo  {journal} {Phys.
  Rev. B}\ }\textbf {\bibinfo {volume} {77}},\ \bibinfo {pages} {094421}
  (\bibinfo {year} {2008})}\BibitemShut {NoStop}%
\bibitem [{\citenamefont {Kaneko}\ \emph {et~al.}(2006)\citenamefont {Kaneko},
  \citenamefont {Metoki}, \citenamefont {Matsuda},\ and\ \citenamefont
  {Kohgi}}]{Kaneko2006}%
  \BibitemOpen
  \bibfield  {author} {\bibinfo {author} {\bibfnamefont {K.}~\bibnamefont
  {Kaneko}}, \bibinfo {author} {\bibfnamefont {N.}~\bibnamefont {Metoki}},
  \bibinfo {author} {\bibfnamefont {T.~D.}\ \bibnamefont {Matsuda}}, \ and\
  \bibinfo {author} {\bibfnamefont {M.}~\bibnamefont {Kohgi}},\ }\href
  {\doibase 10.1143/jpsj.75.034701} {\bibfield  {journal} {\bibinfo  {journal}
  {Journal of the Physical Society of Japan}\ }\textbf {\bibinfo {volume}
  {75}},\ \bibinfo {pages} {034701} (\bibinfo {year} {2006})}\BibitemShut
  {NoStop}%
\bibitem [{\citenamefont {Yamaura}\ and\ \citenamefont
  {Hiroi}(2011)}]{Yamaura2011}%
  \BibitemOpen
  \bibfield  {author} {\bibinfo {author} {\bibfnamefont {J.}~\bibnamefont
  {Yamaura}}\ and\ \bibinfo {author} {\bibfnamefont {Z.}~\bibnamefont
  {Hiroi}},\ }\href {\doibase 10.1143/JPSJ.80.054601} {\bibfield  {journal}
  {\bibinfo  {journal} {Journal of the Physical Society of Japan}\ }\textbf
  {\bibinfo {volume} {80}},\ \bibinfo {pages} {054601} (\bibinfo {year}
  {2011})} \BibitemShut {NoStop}%
\bibitem [{\citenamefont {Christensen}\ \emph {et~al.}(2008)\citenamefont
  {Christensen}, \citenamefont {Abrahamsen}, \citenamefont {Christensen},
  \citenamefont {Juranyi}, \citenamefont {Andersen}, \citenamefont {Lefmann},
  \citenamefont {Andreasson}, \citenamefont {Bahl},\ and\ \citenamefont
  {Iversen}}]{Christensen2008}%
  \BibitemOpen
  \bibfield  {author} {\bibinfo {author} {\bibfnamefont {M.}~\bibnamefont
  {Christensen}}, \bibinfo {author} {\bibfnamefont {A.~B.}\ \bibnamefont
  {Abrahamsen}}, \bibinfo {author} {\bibfnamefont {N.~B.}\ \bibnamefont
  {Christensen}}, \bibinfo {author} {\bibfnamefont {F.}~\bibnamefont
  {Juranyi}}, \bibinfo {author} {\bibfnamefont {N.~H.}\ \bibnamefont
  {Andersen}}, \bibinfo {author} {\bibfnamefont {K.}~\bibnamefont {Lefmann}},
  \bibinfo {author} {\bibfnamefont {J.}~\bibnamefont {Andreasson}}, \bibinfo
  {author} {\bibfnamefont {C.~R.~H.}\ \bibnamefont {Bahl}}, \ and\ \bibinfo
  {author} {\bibfnamefont {B.~B.}\ \bibnamefont {Iversen}},\ }\href
  {http://dx.doi.org/10.1038/nmat2273} {\bibfield  {journal} {\bibinfo
  {journal} {Nature Materials}\ }\textbf {\bibinfo {volume} {7}},\ \bibinfo
  {pages} {811 EP } (\bibinfo {year} {2008})}\BibitemShut {NoStop}%
\bibitem [{\citenamefont {Smith}\ \emph {et~al.}(1985)\citenamefont {Smith},
  \citenamefont {Dolling}, \citenamefont {Kunii}, \citenamefont {Kasaya},
  \citenamefont {Liu}, \citenamefont {Takegahara}, \citenamefont {Kasuya},\
  and\ \citenamefont {Goto}}]{Smith1985}%
  \BibitemOpen
  \bibfield  {author} {\bibinfo {author} {\bibfnamefont {H.~G.}\ \bibnamefont
  {Smith}}, \bibinfo {author} {\bibfnamefont {G.}~\bibnamefont {Dolling}},
  \bibinfo {author} {\bibfnamefont {S.}~\bibnamefont {Kunii}}, \bibinfo
  {author} {\bibfnamefont {M.}~\bibnamefont {Kasaya}}, \bibinfo {author}
  {\bibfnamefont {B.}~\bibnamefont {Liu}}, \bibinfo {author} {\bibfnamefont
  {K.}~\bibnamefont {Takegahara}}, \bibinfo {author} {\bibfnamefont
  {T.}~\bibnamefont {Kasuya}}, \ and\ \bibinfo {author} {\bibfnamefont
  {T.}~\bibnamefont {Goto}},\ }\href {\doibase
  http://dx.doi.org/10.1016/0038-1098(85)90674-X} {\bibfield  {journal}
  {\bibinfo  {journal} {Solid State Communications}\ }\textbf {\bibinfo
  {volume} {53}},\ \bibinfo {pages} {15} (\bibinfo {year} {1985})}\BibitemShut
  {NoStop}%
\bibitem [{\citenamefont {Kunii}\ \emph {et~al.}(1997)\citenamefont {Kunii},
  \citenamefont {Effantin},\ and\ \citenamefont {Rossat-Mingnod}}]{Kunii1997}%
  \BibitemOpen
  \bibfield  {author} {\bibinfo {author} {\bibfnamefont {S.}~\bibnamefont
  {Kunii}}, \bibinfo {author} {\bibfnamefont {J.~M.}\ \bibnamefont {Effantin}},
  \ and\ \bibinfo {author} {\bibfnamefont {J.}~\bibnamefont {Rossat-Mingnod}},\
  }\href {\doibase 10.1143/JPSJ.66.1029} {\bibfield  {journal} {\bibinfo
  {journal} {Journal of the Physical Society of Japan}\ }\textbf {\bibinfo
  {volume} {66}},\ \bibinfo {pages} {1029} (\bibinfo {year} {1997})}\BibitemShut {NoStop}%
\bibitem [{\citenamefont {Lee}\ \emph {et~al.}(2006)\citenamefont {Lee},
  \citenamefont {Hase}, \citenamefont {Sugawara}, \citenamefont {Yoshizawa},\
  and\ \citenamefont {Sato}}]{Lee2006}%
  \BibitemOpen
  \bibfield  {author} {\bibinfo {author} {\bibfnamefont {C.~H.}\ \bibnamefont
  {Lee}}, \bibinfo {author} {\bibfnamefont {I.}~\bibnamefont {Hase}}, \bibinfo
  {author} {\bibfnamefont {H.}~\bibnamefont {Sugawara}}, \bibinfo {author}
  {\bibfnamefont {H.}~\bibnamefont {Yoshizawa}}, \ and\ \bibinfo {author}
  {\bibfnamefont {H.}~\bibnamefont {Sato}},\ }\href {\doibase
  10.1143/JPSJ.75.123602} {\bibfield  {journal} {\bibinfo  {journal} {Journal
  of the Physical Society of Japan}\ }\textbf {\bibinfo {volume} {75}},\
  \bibinfo {pages} {123602} (\bibinfo {year} {2006})}\BibitemShut {NoStop}%
\bibitem [{\citenamefont {Viennois}\ \emph {et~al.}(2004)\citenamefont
  {Viennois}, \citenamefont {Girard}, \citenamefont {Ravot}, \citenamefont
  {Mutka}, \citenamefont {Koza}, \citenamefont {Terki}, \citenamefont
  {Charar},\ and\ \citenamefont {Tedenac}}]{Viennois2004}%
  \BibitemOpen
  \bibfield  {author} {\bibinfo {author} {\bibfnamefont {R.}~\bibnamefont
  {Viennois}}, \bibinfo {author} {\bibfnamefont {L.}~\bibnamefont {Girard}},
  \bibinfo {author} {\bibfnamefont {D.}~\bibnamefont {Ravot}}, \bibinfo
  {author} {\bibfnamefont {H.}~\bibnamefont {Mutka}}, \bibinfo {author}
  {\bibfnamefont {M.}~\bibnamefont {Koza}}, \bibinfo {author} {\bibfnamefont
  {F.}~\bibnamefont {Terki}}, \bibinfo {author} {\bibfnamefont
  {S.}~\bibnamefont {Charar}}, \ and\ \bibinfo {author} {\bibfnamefont
  {J.}~\bibnamefont {Tedenac}},\ }\href {\doibase
  http://dx.doi.org/10.1016/j.physb.2004.03.107} {\bibfield  {journal}
  {\bibinfo  {journal} {Physica B: Condensed Matter}\ }\textbf {\bibinfo
  {volume} {350}},\ \bibinfo {pages} {E403 } (\bibinfo {year} {2004})},\
  \bibinfo {note} {Proceedings of the Third European Conference on
  Neutron Scattering}\BibitemShut {NoStop}%
\bibitem [{\citenamefont {Iwasa}\ \emph {et~al.}(2007)\citenamefont {Iwasa},
  \citenamefont {Mori}, \citenamefont {Hao}, \citenamefont {Murakami},
  \citenamefont {Kohgi}, \citenamefont {Sugawara},\ and\ \citenamefont
  {Sato}}]{Iwasa2007}%
  \BibitemOpen
  \bibfield  {author} {\bibinfo {author} {\bibfnamefont {K.}~\bibnamefont
  {Iwasa}}, \bibinfo {author} {\bibfnamefont {Y.}~\bibnamefont {Mori}},
  \bibinfo {author} {\bibfnamefont {L.}~\bibnamefont {Hao}}, \bibinfo {author}
  {\bibfnamefont {Y.}~\bibnamefont {Murakami}}, \bibinfo {author}
  {\bibfnamefont {M.}~\bibnamefont {Kohgi}}, \bibinfo {author} {\bibfnamefont
  {H.}~\bibnamefont {Sugawara}}, \ and\ \bibinfo {author} {\bibfnamefont
  {H.}~\bibnamefont {Sato}},\ }\href
  {http://stacks.iop.org/1742-6596/92/i=1/a=012122} {\bibfield  {journal}
  {\bibinfo  {journal} {Journal of Physics: Conference Series}\ }\textbf
  {\bibinfo {volume} {92}},\ \bibinfo {pages} {012122} (\bibinfo {year}
  {2007})}\BibitemShut {NoStop}%
\bibitem [{\citenamefont {Kohgi}\ \emph {et~al.}(2006)\citenamefont {Kohgi},
  \citenamefont {Kuwahara}, \citenamefont {Ogita}, \citenamefont {Udagawa},\
  and\ \citenamefont {Iga}}]{Kohgi2006}%
  \BibitemOpen
  \bibfield  {author} {\bibinfo {author} {\bibfnamefont {M.}~\bibnamefont
  {Kohgi}}, \bibinfo {author} {\bibfnamefont {K.}~\bibnamefont {Kuwahara}},
  \bibinfo {author} {\bibfnamefont {N.}~\bibnamefont {Ogita}}, \bibinfo
  {author} {\bibfnamefont {M.}~\bibnamefont {Udagawa}}, \ and\ \bibinfo
  {author} {\bibfnamefont {F.}~\bibnamefont {Iga}},\ }\href@noop {} {\bibfield
  {journal} {\bibinfo  {journal} {Journal of the Physical Society of Japan}\
  }\textbf {\bibinfo {volume} {75}},\ \bibinfo {pages} {085003/1 } (\bibinfo
  {year} {2006})}\BibitemShut {NoStop}%
\bibitem [{\citenamefont {Takegahara}\ \emph {et~al.}(2001)\citenamefont
  {Takegahara}, \citenamefont {Harima},\ and\ \citenamefont
  {Yanase}}]{Takegahara2001}%
  \BibitemOpen
  \bibfield  {author} {\bibinfo {author} {\bibfnamefont {K.}~\bibnamefont
  {Takegahara}}, \bibinfo {author} {\bibfnamefont {H.}~\bibnamefont {Harima}},
  \ and\ \bibinfo {author} {\bibfnamefont {A.}~\bibnamefont {Yanase}},\ }\href
  {\doibase 10.1143/JPSJ.70.1190} {\bibfield  {journal} {\bibinfo  {journal}
  {Journal of the Physical Society of Japan}\ }\textbf {\bibinfo {volume}
  {70}},\ \bibinfo {pages} {1190} (\bibinfo {year} {2001})}\BibitemShut {NoStop}%
\bibitem [{\citenamefont {Lea}\ \emph {et~al.}(1962)\citenamefont {Lea},
  \citenamefont {Leask},\ and\ \citenamefont {Wolf}}]{LLWolf1962}%
  \BibitemOpen
  \bibfield  {author} {\bibinfo {author} {\bibfnamefont {K.}~\bibnamefont
  {Lea}}, \bibinfo {author} {\bibfnamefont {M.}~\bibnamefont {Leask}}, \ and\
  \bibinfo {author} {\bibfnamefont {W.}~\bibnamefont {Wolf}},\ }\href@noop {}
  {\bibfield  {journal} {\bibinfo  {journal} {J. Phys. Chem. Solids}\ }\textbf
  {\bibinfo {volume} {23}},\ \bibinfo {pages} {1381} (\bibinfo {year}
  {1962})}\BibitemShut {NoStop}%
\bibitem [{\citenamefont {Mullen}\ \emph {et~al.}(1974)\citenamefont {Mullen},
  \citenamefont {Luthi}, \citenamefont {Wang}, \citenamefont {Bucher},
  \citenamefont {Longinotti}, \citenamefont {Maita},\ and\ \citenamefont
  {Ott}}]{Mullen1974}%
  \BibitemOpen
  \bibfield  {author} {\bibinfo {author} {\bibfnamefont {M.~E.}\ \bibnamefont
  {Mullen}}, \bibinfo {author} {\bibfnamefont {B.}~\bibnamefont {Luthi}},
  \bibinfo {author} {\bibfnamefont {P.~S.}\ \bibnamefont {Wang}}, \bibinfo
  {author} {\bibfnamefont {E.}~\bibnamefont {Bucher}}, \bibinfo {author}
  {\bibfnamefont {L.~D.}\ \bibnamefont {Longinotti}}, \bibinfo {author}
  {\bibfnamefont {J.~P.}\ \bibnamefont {Maita}}, \ and\ \bibinfo {author}
  {\bibfnamefont {H.~R.}\ \bibnamefont {Ott}},\ }\href {\doibase
  10.1103/PhysRevB.10.186} {\bibfield  {journal} {\bibinfo  {journal} {Phys.
  Rev. B}\ }\textbf {\bibinfo {volume} {10}},\ \bibinfo {pages} {186} (\bibinfo
  {year} {1974})}\BibitemShut {NoStop}%
\bibitem [{\citenamefont {Morin}\ and\ \citenamefont
  {Schmitt}(1990)}]{MorinSchmitt1990}%
  \BibitemOpen
  \bibfield  {author} {\bibinfo {author} {\bibfnamefont {P.}~\bibnamefont
  {Morin}}\ and\ \bibinfo {author} {\bibfnamefont {D.}~\bibnamefont
  {Schmitt}},\ }\enquote {\bibinfo {title} {Quadrupolar interactions and
  magnetoelastic effects in rare earth intermetallic compounds},}\ \ (\bibinfo
  {publisher} {Elsevier Science},\ \bibinfo {year} {1990})\ Chap.~\bibinfo
  {chapter} {1}, pp.\ \bibinfo {pages} {1--132}\BibitemShut {NoStop}%
\bibitem [{\citenamefont {Stevens}(1952)}]{Stevens1952}%
  \BibitemOpen
  \bibfield  {author} {\bibinfo {author} {\bibfnamefont {K.}~\bibnamefont
  {Stevens}},\ }\href@noop {} {\bibfield  {journal} {\bibinfo  {journal} {Proc.
  Phys. Soc. (London)}\ }\textbf {\bibinfo {volume} {A 65}},\ \bibinfo {pages}
  {209} (\bibinfo {year} {1952})}\BibitemShut {NoStop}%
\bibitem [{\citenamefont {Gal{\'e}ra}\ \emph {et~al.}(2015)\citenamefont
  {Gal{\'e}ra}, \citenamefont {Opagiste}, \citenamefont {Amara}, \citenamefont
  {Zbiri},\ and\ \citenamefont {Rols}}]{Galera2015}%
  \BibitemOpen
  \bibfield  {author} {\bibinfo {author} {\bibfnamefont {R.~M.}\ \bibnamefont
  {Gal{\'e}ra}}, \bibinfo {author} {\bibfnamefont {C.}~\bibnamefont
  {Opagiste}}, \bibinfo {author} {\bibfnamefont {M.}~\bibnamefont {Amara}},
  \bibinfo {author} {\bibfnamefont {M.}~\bibnamefont {Zbiri}}, \ and\ \bibinfo
  {author} {\bibfnamefont {S.}~\bibnamefont {Rols}},\ }\href
  {http://stacks.iop.org/1742-6596/592/i=1/a=012011} {\bibfield  {journal}
  {\bibinfo  {journal} {Journal of Physics: Conference Series}\ }\textbf
  {\bibinfo {volume} {592}},\ \bibinfo {pages} {012011} (\bibinfo {year}
  {2015})}\BibitemShut {NoStop}%
\bibitem [{\citenamefont {Maple}\ \emph {et~al.}(2002)\citenamefont {Maple},
  \citenamefont {Ho}, \citenamefont {Zapf}, \citenamefont {Frederick},
  \citenamefont {Bauer}, \citenamefont {Yuhasz}, \citenamefont {Woodward},\
  and\ \citenamefont {Lynn}}]{Maple2002}%
  \BibitemOpen
  \bibfield  {author} {\bibinfo {author} {\bibfnamefont {M.~B.}\ \bibnamefont
  {Maple}}, \bibinfo {author} {\bibfnamefont {P.-C.}\ \bibnamefont {Ho}},
  \bibinfo {author} {\bibfnamefont {V.~S.}\ \bibnamefont {Zapf}}, \bibinfo
  {author} {\bibfnamefont {N.~A.}\ \bibnamefont {Frederick}}, \bibinfo {author}
  {\bibfnamefont {E.~D.}\ \bibnamefont {Bauer}}, \bibinfo {author}
  {\bibfnamefont {W.~M.}\ \bibnamefont {Yuhasz}}, \bibinfo {author}
  {\bibfnamefont {F.~M.}\ \bibnamefont {Woodward}}, \ and\ \bibinfo {author}
  {\bibfnamefont {J.~W.}\ \bibnamefont {Lynn}},\ }\href {\doibase
  10.1143/JPSJS.71S.23} {\bibfield  {journal} {\bibinfo  {journal} {Journal of
  the Physical Society of Japan}\ }\textbf {\bibinfo {volume} {71}},\ \bibinfo
  {pages} {23} (\bibinfo {year} {2002})} \BibitemShut {NoStop}%
\bibitem [{\citenamefont {Matsuhira}\ \emph {et~al.}(2005)\citenamefont
  {Matsuhira}, \citenamefont {Doi}, \citenamefont {Wakeshima}, \citenamefont
  {Hinatsu}, \citenamefont {Kihou}, \citenamefont {Sekine},\ and\ \citenamefont
  {Shirotani}}]{Matsuhira2005}%
  \BibitemOpen
  \bibfield  {author} {\bibinfo {author} {\bibfnamefont {K.}~\bibnamefont
  {Matsuhira}}, \bibinfo {author} {\bibfnamefont {Y.}~\bibnamefont {Doi}},
  \bibinfo {author} {\bibfnamefont {M.}~\bibnamefont {Wakeshima}}, \bibinfo
  {author} {\bibfnamefont {Y.}~\bibnamefont {Hinatsu}}, \bibinfo {author}
  {\bibfnamefont {K.}~\bibnamefont {Kihou}}, \bibinfo {author} {\bibfnamefont
  {C.}~\bibnamefont {Sekine}}, \ and\ \bibinfo {author} {\bibfnamefont
  {I.}~\bibnamefont {Shirotani}},\ }\href {\doibase
  https://doi.org/10.1016/j.physb.2005.01.282} {\bibfield  {journal} {\bibinfo
  {journal} {Physica B: Condensed Matter}\ }\textbf {\bibinfo {volume}
  {359-361}},\ \bibinfo {pages} {977 } (\bibinfo {year} {2005})},\ \bibinfo
  {note} {proceedings of the International Conference on Strongly Correlated
  Electron Systems}\BibitemShut {NoStop}%
\bibitem [{\citenamefont {Rudenko}\ \emph {et~al.}(2016)\citenamefont
  {Rudenko}, \citenamefont {Henkie},\ and\ \citenamefont
  {Cichorek}}]{Rudenko2016}%
  \BibitemOpen
  \bibfield  {author} {\bibinfo {author} {\bibfnamefont {A.}~\bibnamefont
  {Rudenko}}, \bibinfo {author} {\bibfnamefont {Z.}~\bibnamefont {Henkie}}, \
  and\ \bibinfo {author} {\bibfnamefont {T.}~\bibnamefont {Cichorek}},\ }\href
  {\doibase https://doi.org/10.1016/j.ssc.2016.04.026} {\bibfield  {journal}
  {\bibinfo  {journal} {Solid State Communications}\ }\textbf {\bibinfo
  {volume} {242}},\ \bibinfo {pages} {21 } (\bibinfo {year}
  {2016})}\BibitemShut {NoStop}%
\bibitem [{\citenamefont {Amara}\ \emph {et~al.}(2005)\citenamefont {Amara},
  \citenamefont {Luca}, \citenamefont {Gal{\'e}ra}, \citenamefont {Givord},
  \citenamefont {Detlefs},\ and\ \citenamefont {Kunii}}]{Amara2005}%
  \BibitemOpen
  \bibfield  {author} {\bibinfo {author} {\bibfnamefont {M.}~\bibnamefont
  {Amara}}, \bibinfo {author} {\bibfnamefont {S.}~\bibnamefont {Luca}},
  \bibinfo {author} {\bibfnamefont {R.-M.}\ \bibnamefont {Gal{\'e}ra}},
  \bibinfo {author} {\bibfnamefont {F.}~\bibnamefont {Givord}}, \bibinfo
  {author} {\bibfnamefont {C.}~\bibnamefont {Detlefs}}, \ and\ \bibinfo
  {author} {\bibfnamefont {S.}~\bibnamefont {Kunii}},\ }\href@noop {}
  {\bibfield  {journal} {\bibinfo  {journal} {Phys. Rev. B}\ }\textbf {\bibinfo
  {volume} {72}},\ \bibinfo {pages} {64447} (\bibinfo {year}
  {2005})}\BibitemShut {NoStop}%
\bibitem [{\citenamefont {Amara}\ \emph {et~al.}(2010)\citenamefont {Amara},
  \citenamefont {Gal\'era}, \citenamefont {Aviani},\ and\ \citenamefont
  {Givord}}]{Amara2010}%
  \BibitemOpen
  \bibfield  {author} {\bibinfo {author} {\bibfnamefont {M.}~\bibnamefont
  {Amara}}, \bibinfo {author} {\bibfnamefont {R.-M.}\ \bibnamefont {Gal\'era}},
  \bibinfo {author} {\bibfnamefont {I.}~\bibnamefont {Aviani}}, \ and\ \bibinfo
  {author} {\bibfnamefont {F.}~\bibnamefont {Givord}},\ }\href {\doibase
  10.1103/PhysRevB.82.224411} {\bibfield  {journal} {\bibinfo  {journal} {Phys.
  Rev. B}\ }\textbf {\bibinfo {volume} {82}},\ \bibinfo {pages} {224411}
  (\bibinfo {year} {2010})}\BibitemShut {NoStop}%
\bibitem [{\citenamefont {Sirota}\ \emph {et~al.}(2000)\citenamefont {Sirota},
  \citenamefont {Novikov},\ and\ \citenamefont {Novikov}}]{Sirota2000}%
  \BibitemOpen
  \bibfield  {author} {\bibinfo {author} {\bibfnamefont {N.~N.}\ \bibnamefont
  {Sirota}}, \bibinfo {author} {\bibfnamefont {V.~V.}\ \bibnamefont {Novikov}},
  \ and\ \bibinfo {author} {\bibfnamefont {A.~V.}\ \bibnamefont {Novikov}},\
  }\href {\doibase 10.1134/1.1324045} {\bibfield  {journal} {\bibinfo
  {journal} {Physics of the Solid State}\ }\textbf {\bibinfo {volume} {42}},\
  \bibinfo {pages} {2093} (\bibinfo {year} {2000})}\BibitemShut {NoStop}%
\bibitem [{\citenamefont {Iwasa}\ \emph {et~al.}(2006)\citenamefont {Iwasa},
  \citenamefont {Kohgi}, \citenamefont {Sugawara},\ and\ \citenamefont
  {Sato}}]{IWASA2006}%
  \BibitemOpen
  \bibfield  {author} {\bibinfo {author} {\bibfnamefont {K.}~\bibnamefont
  {Iwasa}}, \bibinfo {author} {\bibfnamefont {M.}~\bibnamefont {Kohgi}},
  \bibinfo {author} {\bibfnamefont {H.}~\bibnamefont {Sugawara}}, \ and\
  \bibinfo {author} {\bibfnamefont {H.}~\bibnamefont {Sato}},\ }\href {\doibase
  https://doi.org/10.1016/j.physb.2006.01.073} {\bibfield  {journal} {\bibinfo
  {journal} {Physica B: Condensed Matter}\ }\textbf {\bibinfo {volume} {378}},\
  \bibinfo {pages} {194 } (\bibinfo {year} {2006})},\ \bibinfo {note}
  {proceedings of the International Conference on Strongly Correlated Electron
  Systems}\BibitemShut {NoStop}%
\bibitem [{\citenamefont {Callen}\ and\ \citenamefont
  {Callen}(1963)}]{Callen1963}%
  \BibitemOpen
  \bibfield  {author} {\bibinfo {author} {\bibfnamefont {E.~R.}\ \bibnamefont
  {Callen}}\ and\ \bibinfo {author} {\bibfnamefont {H.~B.}\ \bibnamefont
  {Callen}},\ }\href {\doibase 10.1103/PhysRev.129.578} {\bibfield  {journal}
  {\bibinfo  {journal} {Phys. Rev.}\ }\textbf {\bibinfo {volume} {129}},\
  \bibinfo {pages} {578} (\bibinfo {year} {1963})}\BibitemShut {NoStop}%
\end{thebibliography}
\end{document}